\documentclass[onecolumn]{gji}
\let\bibhang\relax
\usepackage{natbib}
\usepackage{timet}
\usepackage{amsmath}

\usepackage{graphicx}
\usepackage{amssymb}
\usepackage{color}
\newcommand{\bx}{{\bf x}}
\newcommand{\measur}{{\mathcal F}}
\newcommand{\bnabla}{{\boldsymbol{\nabla}}}

\newcommand{\cc}{{\mathcal C}}

\title[Seismic noise cross-correlation amplitudes]
  {The influence of noise sources on cross-correlation amplitudes}
\author[Shravan M. Hanasoge]
  {Shravan M. Hanasoge\thanks{hanasoge@princeton.edu} \\
  Department of Geosciences, Princeton University, NJ 08544, USA\\
  Max-Planck-Institut f\"{u}r Sonnensystemforschung, 37191 Katlenburg-Lindau, Germany
  }
\date{Received 2012 June 18; in original form 2012 April 12}

\begin{document}

\label{firstpage}

\maketitle

\begin{summary}
We use analytical examples and asymptotic forms to examine the mathematical structure
and physical meaning of the seismic cross correlation measurement.
We show that in general, cross correlations are {\it not} Green's functions of medium, and may be very different
depending on the source distribution. 
The modeling of noise sources using spatial distributions as opposed to discrete collections of sources
is emphasized. 
{When stations are illuminated by spatially complex source distributions, cross correlations show arrivals at
a variety of time lags, from zero to the maximum surface-wave arrival time.
Here, we demonstrate the possibility of inverting for the source distribution using the energy of the full cross-correlation waveform.} 
The interplay between the source distribution and wave attenuation in determining the functional dependence of cross correlation energies
on station-pair distance is quantified. Without question, energies contain information about wave attenuation. However,
the accurate interpretation of such measurements is tightly connected to the knowledge of the source distribution.
\end{summary}

\section{Introduction}
Terrestrial seismic noise is generated at a range of temporal frequencies, by human activity,
storms, oceanic wave microseisms \cite[e.g.,][]{longuet50, kedar05, stehly06} and the 
ocean-excited low-frequency hum of Earth \cite[e.g.,][]{nawa98, rhie04}. Seismic noise is as used as a compelling alternative
to earthquake tomography to image the crust. Most importantly, it enables the study of temporal variations of the crust \cite[e.g.,][]{wegler07,BrenguierSCI08,shapiro11, rivet11}
and volcanoes \cite[e.g.,][]{campillo07}. The 
cross correlation measurement has a physical flavor that is intrinsically different from the classical
tomographic analog, i.e., wavefield displacement. In particular, the time variable in classical tomography is the propagation 
delay between the source and the station whereas the time lag in cross correlation tomography is 
connected to the path difference between the source and the two stations.


%
Under controlled circumstances,
such as a when the source distribution is uniform, representation theorems \cite[e.g.,][]{snieder10} allow for the cross correlation to be
written as a modulation of Green's function between the stations. In other words, such theorems state that an equally weighted sum 
over Green's functions between every source (over all space; constant amplitude) and the stations is equivalent to Green's function
between the stations.
However, Earth noise is typically
anisotropic and in such a scenario, Green's functions along some source-station paths are weighted more strongly than others and the elegant correspondence
may be lost. 
Further, the climate is in continuous flux, and the manner of excitation of seismic noise by, e.g., ocean waves, changes through
the year \cite[e.g.,][]{stehly06}. Seismology is a precision science and consequently, modeling the source distribution and its effect on the cross correlation 
is critical.

The study of terrestrial seismic noise has strong connections with the seismic wavefields
of stars, and in particular, the Sun. The use of cross correlations of the wavefield of the Sun to probe interior solar structure
was pioneered by \citet{duvall} in a landmark paper. 
The formal interpretation of these measurements had to wait till the advance by \citet{woodard}, who 
laid the theory of cross correlation on a formal statistical foundation. A number of years later, \citet{gizon02}, based on this
work, were able to compute kernels for cross correlations of helioseismic noise arising from a distribution of sources. However, \citet{gizon02} did not account for 3-D heterogeneous
backgrounds and the rewriting of their formalism in the language of adjoint methods was led by \citet{tromp10} for the terrestrial case
and by \citet{hanasoge11} for the helioseismic scenario. One useful concept that emerged from these articles is that of dealing 
with source distributions as opposed to a discrete number of them \cite[e.g.,][]{larose06, tsai09}. More importantly, the results of \citet{tromp10} enabled the
computational prediction (in a forward sense) of cross correlations based on a given Earth model and source distribution 
\citep[this problem has received considerable attention: e.g.,][]{hellegji06,ChevrotJGR07,YangG308,WeaverJASA09,cupillard10,tsai10,FromentGEO10}.

In this article, we discuss some of the concepts underlying the cross correlation measurement using a simple 2-D example. Section~\ref{formal}
deals with the cross correlation and its connection to the source distribution. Storms, which excite seismic waves, but sometimes physically move substantial 
distances over a span of days (i.e., over the measurement window), can be modeled as well, albeit  through a more complex ansatz for the distribution.
We also introduce the basic partial differential equation governing the 
wavefield and Green's function for the simple case studied here. The analysis of the variations of the cross correlation due to the changes
in the source distribution, thereby leading to the sensitivity kernel are discussed in Section~\ref{struct.source}. In particular, its asymptotic form
reveals the structure of the source-amplitude kernel. The cross-correlation energy misfit and its kernel are discussed and computed semi-analytically
in Section~\ref{compute.kern}. Based on this formalism, the impact of non-uniform source distributions on cross correlations
is examined and we make a case for imaging of the source distribution and briefly discuss the limitations in Section~\ref{nonuni.dist}. 

The operator formulation of the adjoint method discussed by, e.g., \citet{fichtner06}, which works elegantly for classical tomography, unfortunately 
does not naturally apply to higher order measurements. In classical tomography,
we vary the wavefield, which is directly the solution to the wave operator. To create an equivalent operator formalism for cross-correlation tomography, one
needs to write a differential equation for the cross correlation itself, which is impractical. Consequently, we must
carry out the Born expansion by brute force and analyze the resultant terms \citep[e.g.,][]{tromp10, hanasoge11}, as discussed in Section~\ref{scatter} of this article. 
Three Green's functions appear in this expansion (as opposed to two in the classical tomography case) and their role in modeling scattering is elucidated.
The source distribution plays a critical role in determining cross-correlation energies. In order to accurately interpret cross-correlation energies in the context of
wave attenuation or scattering, the source distribution must be well known, as discussed in Section~\ref{attenuation}. Indeed, once the effect of sources has been
accounted for, cross-correlation energies contain information about attenuation. In a dense network, the sensitivity to attenuation and scattering is primarily restricted to the region
within the network because the hyperbolic features that appear in cross-correlation kernels for attenuation \citep[e.g., see Figure 6 of ][]{tromp10} cancel.
We conclude in Section~\ref{conclude}.



\section{Formal interpretation}\label{formal}
Cross correlations of seismic noise fluctuations $\phi(\bx, t)$, denoted by $\cc_{\alpha\beta}(t)$, are defined as
\begin{equation}
\cc_{\alpha\beta}(t;T) = \int_0^T\,dt'\,\phi(\bx_\alpha,t')\,\phi(\bx_\beta,t'+t),\label{cctime}
\end{equation}
where $T$ is the temporal length of averaging, $t$ is time and $\bx_\alpha,\bx_\beta$ are spatial locations
at which measurements are made. In order to obtain cross correlations of a reasonable signal-to-noise ratio, 
$T$ must be on the order of several source correlation times and wave travel times between source-station pairs. 
As the temporal window of averaging $T$ grows, the cross correlation
approaches a limiting value (provided the source distribution and the medium do not change substantially over this time scale),
i.e., what we term the {\it expectation value}. This was labelled the {\it ensemble cross correlation} by \citet{tromp10} in order to describe
ensemble averaging over many source times \citep[and realizations; see also, e.g.,][for convergence studies]{larose08, cupillard10}. 
Moments of stochastic processes for which expectation values exist are termed {\it ergodic}. 
Terrestrial seismic noise is ergodic because wave excitation of oceanic origin appears to have well
behaved statistics (e.g., when the sources are described by a Gaussian random process). 

The relation~(\ref{cross.c})
when applied to equation~(\ref{cctime}) allow us to describe the cross correlation in temporal Fourier domain
\begin{equation}
\cc_{\alpha\beta}(\omega) = \phi^*(\bx_\alpha,\omega)\,\phi(\bx_\beta,\omega).\label{correl}
\end{equation}
Denoting the limit (or expected) cross correlation by $\langle\cc_{\alpha\beta}(\omega)\rangle$, we have
\begin{equation}
\langle\cc_{\alpha\beta}\rangle = \langle\phi^*(\bx_\alpha,\omega)\,\phi(\bx_\beta,\omega)\rangle.\label{expect}
\end{equation}

\begin{figure}
\centering
\includegraphics*[width=\linewidth]{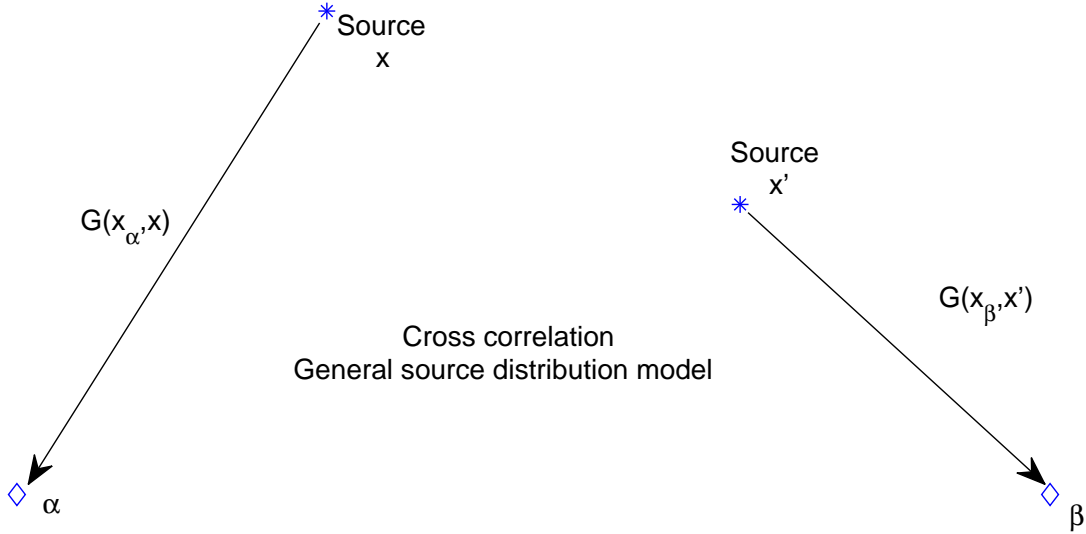}
\caption{Cross correlations from a general model of sources as stated in equation~(\ref{ccgen}). The source at $\bx$ excites waves that propagate to $\alpha$ and 
the source at ${\bx}'$ generates waves that propagate to $\beta$. Because the sources themselves are statistically correlated, the expectation value of the wavefield cross correlation
measured at points $\bx_\alpha, \bx_\beta$ is non-trivial. Note that if the sources at two spatial points were statistically independent, the expectation value of
the cross correlation would be zero when $\bx \neq \bx'$, as shown in Figure~\ref{cc_here}. This form of source distribution is useful in modeling, e.g.,
storms, which can move substantial distances over relatively short times.\label{cc_general_sources}}
\end{figure}

\begin{figure}
\centering
\includegraphics*[width=\linewidth]{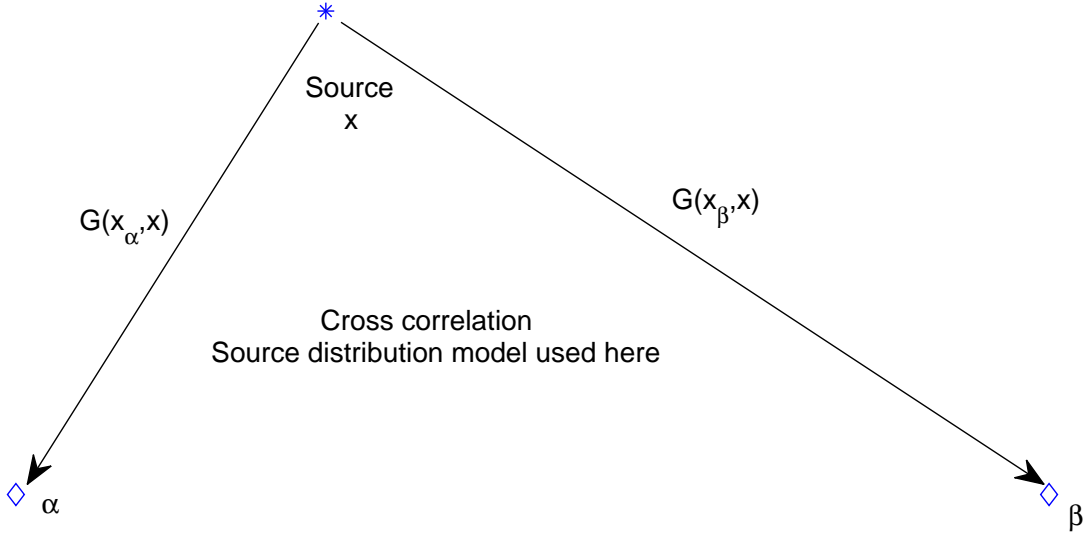}
\caption{Cross correlations due to the distribution with $\langle S^*(\bx,\omega)\,S(\bx',\omega)\rangle = {\mathcal P}(\omega)\,\sigma(\bx)\,\delta(\bx-\bx')$, as described in
equation~(\ref{cc.eq}).
This model renders feasible the prediction or forward computation of the cross correlations because of the reduction in the number of integration variables. The
inverse problem is also more easily dealt with since the number of degrees of freedom is much smaller.\label{cc_here}}
\end{figure}

The wave equation we consider here is
\begin{equation}
\rho\partial_t^2\phi - \bnabla\cdot(c^2\bnabla\phi) = S(\bx,t),\label{2deq}
\end{equation}
where $\rho$ is density, $\bx= (x,y)$ is a 2-D flat space, $t$ time,
 $\phi$ the wave displacement, $\bnabla =( \partial_x, \partial_y)$ the covariant spatial derivative, $S(\bx,t)$ the
 source and $c$ wavespeed. For the simple case considered here, we assume constant wavespeed $c$. 
 Green's function $G(\bx,\bx';t)$ for the displacement at $(\bx,t)$ due to a spatio-temporal delta source at $(\bx',0)$ is 
the solution to
 \begin{equation}
(\rho \partial^2_t - c^2\nabla^2)G(\bx,\bx';t) = \delta(\bx-\bx')\delta(t).
 \end{equation}
This equation is explicitly solvable; Green's function in temporal Fourier-transform 
(according to the convention defined in appendix~\ref{convention}) is given by \citep[e.g.,][]{aki80}
 \begin{equation}
 G(\bx,\bx',\omega) = H_0^{(1)}\left(\frac{\omega}{c}|\bx-\bx'|\right),\label{green.hank}
 \end{equation}
 where $\omega$ is temporal frequency and $H_0^{(1)}$ is the Hankel function of the first kind.
 {This is also approximately the surface wave portion of Green's function for a laterally homogeneous Earth}.
 To lighten notational burden, we cease to explicitly state frequency $\omega$ unless
 required. The wavefield $\phi(\bx)$ excited by sources $S(\bx')$ is described by
 \begin{equation}
 \phi(\bx) = \int d\bx'\,G(\bx, \bx')\,S(\bx').\label{green.eq}
 \end{equation}
The correlation in Fourier domain~(\ref{correl}) may be rewritten in terms of Green's functions
and sources
\begin{equation}
\cc_{\alpha\beta}(\omega) = \int d\bx'\int d\bx\,G^*(\bx_\alpha,\bx)\,G(\bx_\beta,\bx')\,S^*(\bx)\,S(\bx'),\label{exp.eq}
\end{equation}
and the expected cross correlation~(\ref{expect}) becomes
\begin{equation}
\langle\cc_{\alpha\beta}\rangle =\int d\bx'\int d\bx\,G^*(\bx_\alpha,\bx)\,G(\bx_\beta,\bx')\,\langle S^*(\bx)\,S(\bx')\rangle,\label{ccgen}
\end{equation}
where the ensemble averaging has been brought into the integral and placed around the source terms. This is the
point at which we have moved from treating dynamically evolving sources to studying their statistics. Thus we have taken a
system whose source distribution is unknown and posed it in terms of a (potentially) computable statistical theory. 
Equation~(\ref{exp.eq}) states that a wave excited at $\bx$ propagates, through a medium described by Green's function, to point $\bx_\alpha$ 
and similarly form $\bx'$ to $\bx_\beta$, pictorially depicted in Figure~\ref{cc_general_sources}. 
Contributions from wave sources over all space are summed to produce the wavefield at points $\bx_\alpha, \bx_\beta$,
which explains the spatial integrals.
For a complete
theory, we need to include the statistical spatial covariance of the source distribution, i.e., $\Lambda(\bx,\bx',\omega) =\langle S^*(\bx,\omega)\,S(\bx',\omega)\rangle$,
but such a problem is very hard to study. Consequently, we model spatially uncorrelated sources, i.e., $\Lambda(\bx,\bx',\omega) = {\mathcal P}(\omega)\,\sigma(\bx)\,\delta(\bx-\bx')$,
where ${\mathcal P}$ is the power spectrum and $\sigma(\bx)$ is the source distribution in space. 
This choice greatly reduces the number of degrees of freedom in any eventual inverse problem (see Figure~\ref{cc_here}). Note that the source distribution
typically varies as a function of frequency, and this can be modeled by studying narrowly filtered cross correlations such that we may invert for a different spatial distribution in each frequency window.
Different parametrizations of $\Lambda$ may be chosen depending on the problem at hand.
Using this assumption, we have
\begin{equation}
\langle\cc_{\alpha\beta}\rangle =\int d\bx\,G^*(\bx_\alpha,\bx)\,G(\bx_\beta,\bx)\,{\mathcal P}(\omega)\,\sigma(\bx),\label{cc.eq}
\end{equation}
which allows us to construct forward models of cross correlation, a first step towards inversions.
Substituting Green's function~(\ref{green.hank}) into~(\ref{cc.eq}), we obtain
\begin{equation}
\langle\cc_{\alpha\beta}\rangle =\int d\bx\,H_0^{(2)}\left(\frac{\omega}{c}|\bx_\alpha-\bx|\right)\,H_0^{(1)}\left(\frac{\omega}{c}|\bx_\beta-\bx|\right)\,{\mathcal P}\,\sigma(\bx),\label{cc.hank}
\end{equation}
where the Hankel function of the second kind is defined thus $H_0^{(2)} = H_0^{(1)*}$. For convenience, we also define $\Delta_\alpha = |\bx_\alpha-\bx|$.

\section{Structure of Source Kernels}\label{struct.source}
\begin{figure}
\centering
\includegraphics*[width=\linewidth]{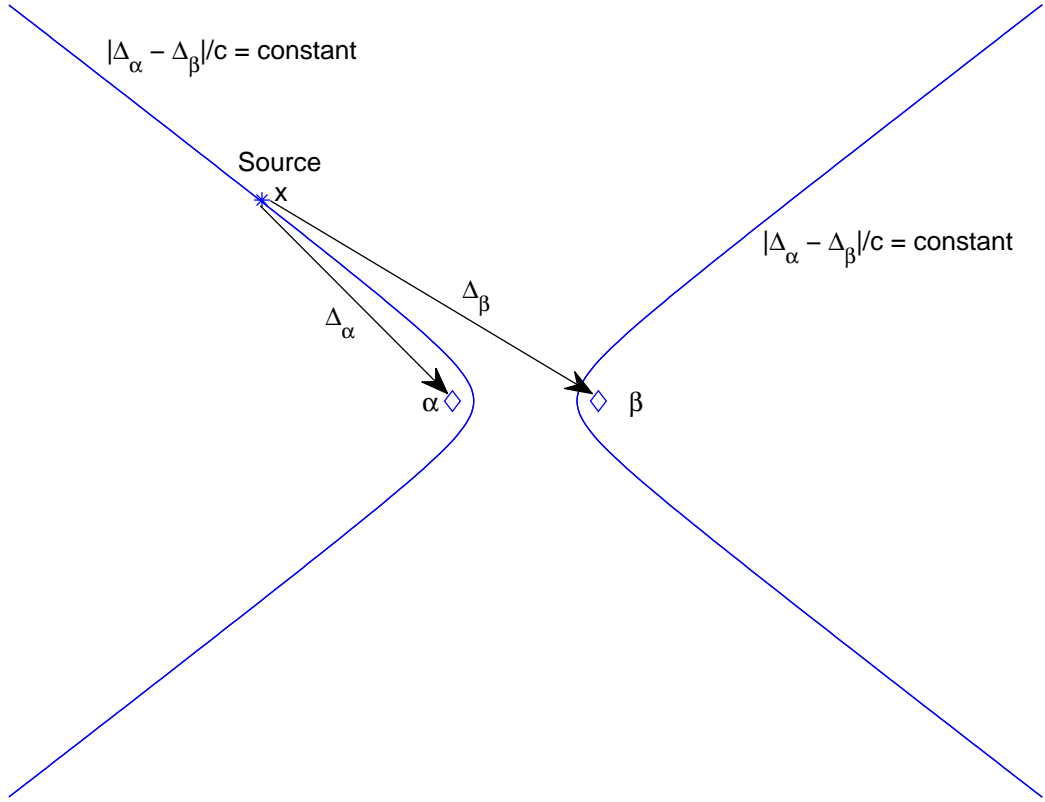}
\caption{Contours of constant path difference are hyperbolae whose foci are the two stations $\bx_\alpha, \bx_\beta$. For a given time-lag measurement of the cross correlation
(i.e., choice of measurement time window), sources along the hyperbola constructively contribute to it. \label{hyperbolae}}
\end{figure}

Sensitivity kernels for noise distributions were introduced by \citet{tromp10}, who termed them {\it ensemble kernels}.
Consider the inverse problem where
we are interested solely in the source distribution, i.e., variations of the correlation are rooted only in variations of the
source distribution (as opposed to a more complete inverse problem which would contain variations to structure as well)
\begin{equation}
\delta\langle\cc_{\alpha\beta}\rangle= \langle\delta\cc_{\alpha\beta}\rangle =\int d\bx\,H_0^{(2)}\left(\frac{\omega}{c}\Delta_\alpha\right)\,H_0^{(1)}\left(\frac{\omega}{c}\Delta_\beta\right)\,
{\mathcal P}\,\delta\sigma(\bx).
\end{equation}
Suppose the measurable $\delta\measur$, such as a travel time or energy, is locally a linear functional of the variation of the cross correlation, i.e.,
\begin{equation}
\delta\measur = \int d\omega\,W^*_{\alpha\beta}\, \langle\delta\cc_{\alpha\beta}\rangle,
\end{equation}
where $W_{\alpha\beta}(\omega)$ is some weight function. Then we have
\begin{equation}
\delta\measur = \int d\bx\left[\int d\omega\,W^*_{\alpha\beta}\,H_0^{(2)}\left(\frac{\omega}{c}\Delta_\alpha\right)\,H_0^{(1)}\left(\frac{\omega}{c}\Delta_\beta\right)\right]{\mathcal P} \,\delta\sigma(\bx),\label{measu}
\end{equation}
where it may be seen that the term within the square brackets is the source kernel $K_{\alpha\beta}(\bx)$
\begin{equation}
K_{\alpha\beta}(\bx) = \int d\omega\,W^*_{\alpha\beta}\,H_0^{(2)}\left(\frac{\omega}{c}\Delta_\alpha\right)\,H_0^{(1)}\left(\frac{\omega}{c}\Delta_\beta\right)\,{\mathcal P},\label{kern.source}
\end{equation}
which is the sensitivity to variations in the amplitude of the source distribution.
With a little help from asymptotics,
we can immediately perceive the structure of this kernel. The far-field approximation to Hankel functions, i.e., for large values of the argument, is
\begin{equation}
H_0^{(1)}(z) \sim \frac{2}{\sqrt{\pi z}}\exp\left(iz - i\frac{\pi}{4}\right), \,\,\,\, |z| \rightarrow\infty.
\end{equation}
Thus, were there to be no source activity at very low temporal frequency, at a distance of several wavelengths away from the measurement locations $\bx_{\alpha, \beta}$,
we may write
\begin{eqnarray}
H_0^{(1)}\left(\frac{\omega}{c}\Delta_\beta\right) &\sim& \sqrt\frac{2c}{\pi \omega\Delta_\beta}\exp\left(\frac{i\omega}{c}\Delta_\beta - i\frac{\pi}{4}\right),\\
H_0^{(2)}\left(\frac{\omega}{c}\Delta_\alpha\right) &\sim& \sqrt\frac{2c}{\pi \omega\Delta_\alpha}\exp\left(-\frac{i\omega}{c}\Delta_\alpha + i\frac{\pi}{4}\right).
\end{eqnarray}
Substituting these asymptotic relations into the expression for the kernel in equation~(\ref{kern.source}), we obtain the far-field limit
\begin{equation}
K_{\alpha\beta}(\bx) \sim \frac{1}{\sqrt{\Delta_\alpha\Delta_\beta}}\,\frac{1}{2\pi}\int d\omega\, {\hat f}(\omega)\,\exp\left[-i\omega\frac{\Delta_\alpha - \Delta_\beta}{c}\right],
\end{equation}
where ${\hat f}(\omega) = 4\,c\,W^*_{\alpha\beta}\,{\mathcal P}\,/\omega$. Comparing this to the definition of the inverse Fourier transform~(\ref{inv.fourier}),
we conclude that 
\begin{equation}
K_{\alpha\beta}(\bx) \sim {f\left(\frac{\Delta_\alpha - \Delta_\beta}{c}\right)}\frac{1}{\sqrt{\Delta_\alpha\Delta_\beta}},\label{source.form}
\end{equation}
and the source-distribution kernel is also a function of the path difference between a given point and the two stations, shown in Figure~\ref{hyperbolae}. This relationship suggests the existence
of hyperbolic features in the kernels, i.e., contours along which $(\Delta_\alpha - \Delta_\beta)/c = (|\bx - \bx_\alpha| - |\bx - \bx_\beta|)/c = C$ is constant
\citep[also see, e.g.,][]{snieder04,roux05,cupillard11}.
{The presence of the term $1/\sqrt{\Delta_\alpha\Delta_\beta}$ results in a much greater sensitivity to regions close to the station and the
path joining the stations. In comparison, the source kernel possesses relatively weak sensitivity to areas away from this line.}
Since we have assumed that the source distribution is spatially uncorrelated, contributions that constructively add to the expected value of
the cross correlation can only be from the same point in the source distribution. In other words, a source excites a waves at $\bx$, which propagate
to $\bx_{\alpha, \beta}$ as described by Green's functions between the two stations and the source location. 
The cross correlation thus registers the travel-time difference between a source point and the two stations, which remains constant along
hyperbolae whose foci are the two stations. 
Source kernels are relatively easy to compute since there are no scattering terms. When including scattering, variations of Green's function
come in to play, and a Born expansion is required to determine the variation of Green's functions.

An interesting analogy to note is that between scattering in the classical tomographic case (banana-doughnut kernels) and the source-amplitude kernel in the  
cross-correlation scenario. We will not go into mathematical detail but in both of these cases, two Green's functions participate in the construction
of the kernels. The sole difference between these two scenarios is that the source-amplitude kernel consists of a {\it correlation} between two Green's functions
and the scattering kernel is composed of a {\it convolution} between Green's functions \citep[e.g.,][]{dahlen99}. The source-amplitude kernel in the cross
correlation case is thus analogous to {\it anti-causal} scattering in the classical case since one of the two Green's functions has a complex conjugate
 in the cross-correlation case. The correlation/convolution difference
changes the character of the kernel: elliptical features are observed in classical banana-doughnut kernels \citep[e.g.,][]{dahlen99} whereas hyperbolic features are
seen in the cross-correlation source-amplitude kernels \citep[also see,][]{gizon02}. Equation~(\ref{kern.source}) for the classical scattering analog would then be
\begin{equation}
K^{\mathrm CS}_{\alpha\beta}(\bx) = \int d\omega\,W^*_{\alpha\beta}\,H_0^{(1)}\left(\frac{\omega}{c}\Delta_\alpha\right)\,H_0^{(1)}\left(\frac{\omega}{c}\Delta_\beta\right),\label{classical.scat}
\end{equation}
where now this represents a convolution (note that both are Hankel functions of the first kind) and where
$K^{\mathrm CS}_{\alpha\beta}(\bx)$ is a classical scattering kernel. Applying the same asymptotic analysis, we obtain
\begin{equation}
K^{\mathrm CS}_{\alpha\beta}(\bx) \sim {f\left(\frac{\Delta_\alpha + \Delta_\beta}{c}\right)}\frac{1}{\sqrt{\Delta_\alpha\Delta_\beta}},\label{scat.form}
\end{equation}
which will produce elliptical features since these are contours of constant path length, i.e., where $(\Delta_\alpha + \Delta_\beta)/c = (|\bx - \bx_\alpha| + |\bx - \bx_\beta|)/c = C$.

\subsection{Computing Kernels}\label{compute.kern}
In this section, we discuss the structure of source-amplitude kernels through the exact computation of equations~(\ref{cc.hank}) and~(\ref{kern.source}). 
The formulation used in this section follows from that of \citet{dahlen02}. We begin the process of computing kernels by
defining a misfit functional $\chi$ in terms of the measured energy anomaly,
\begin{equation}
\chi= \frac{1}{2}\sum_{\alpha,\beta} \left(\ln\frac{A^{\rm obs}_{\alpha\beta}}{A^{\rm syn}_{\alpha\beta}}\right)^2,\label{misf}
\end{equation}
where the energy is defined as
\begin{equation}
A^{\rm syn}_{\alpha\beta} = \sqrt{\frac{\int dt\,\,w(t)\, \langle{\mathcal C_{\alpha\beta}(t)}\rangle^2}{\int dt\,w(t)} } = \sqrt{\frac{1}{2\pi T}\int d\omega\,\langle{\mathcal C}^*_{\alpha\beta}\rangle\, \langle{\mathcal C}_{\alpha\beta}\rangle },\label{syn.amp}
\end{equation}
where $w(t)$ is the windowing function, ${\mathcal C}_{\alpha\beta}$ is
the windowed cross correlation and $T = \int dt\,w(t)$. Note that we use the term {\it energy} interchangeably with {\it amplitude}.
To preserve simplicity, we do not apply frequency filters, although they may be easily included. {In general, although we may
compute sensitivity kernels for other measurements such as travel times, we restrict ourselves here to the cross correlation energy.}
With a little manipulation, not shown here, the variation in misfit is given by
\begin{equation}
\delta\chi= -\sum_{\alpha,\beta} \left(\ln\frac{A^{\rm obs}_{\alpha\beta}}{A^{\rm syn}_{\alpha\beta}}\right) \frac{\delta A^{\rm syn}_{\alpha\beta}}{A^{\rm syn}_{\alpha\beta}} 
= -\sum_{\alpha,\beta}\left(\frac{1}{A^{\rm syn}_{\alpha\beta}}\right)^2 \left(\ln\frac{A^{\rm obs}_{\alpha\beta}}{A^{\rm syn}_{\alpha\beta}}\right)\frac{1}{2\pi T}\int d\omega\,\langle{\mathcal C}_{\alpha\beta}^*\rangle \, \langle\delta{\mathcal C}_{\alpha\beta}\rangle.\label{misfit.var}
\end{equation}
If one were to, as before, assume that the variations in the cross correlation only arose from changes to the source distribution,
\begin{equation}
\langle\delta{\mathcal C}_{\alpha\beta}\rangle = \int d\bx\,G^*(\Delta_{\alpha})\,G(\Delta_\beta)\,{\mathcal P}(\omega)\,\delta\sigma(\bx).
\end{equation}
Thus defining the weight function as 
\begin{equation}
W_{\alpha\beta} = \frac{1}{T}\,\left(\frac{1}{A^{\rm syn}_{\alpha\beta}}\right)^2\langle{{\mathcal C}}_{\alpha\beta}\rangle,\label{weight}
\end{equation}
variations to the misfit functional are then described by
\begin{equation}
\delta\chi = -\sum_{\alpha,\beta} \left(\ln\frac{A^{\rm obs}_{\alpha\beta}}{A^{\rm syn}_{\alpha\beta}}\right) \int d\bx\,K_{\alpha\beta}(\bx)\, \delta \sigma(\bx).
\end{equation}
The kernel normalization is tested by confirming that the following integral is satisfied
\begin{equation}
\int d\bx\, K_{\alpha\beta}(\bx)\,\sigma(\bx) = \left(\frac{1}{A^{\rm syn}_{\alpha\beta}}\right)^2\,\frac{1}{2\pi T}\int d\omega\, {\langle {\mathcal C}}^*_{\alpha\beta}\rangle \int d\bx\,G^*(\Delta_{\alpha})\,G(\Delta_\beta)\,{\mathcal P}(\omega)\,\sigma(\bx) = 1
\end{equation}
(obtained upon applying definition~(\ref{syn.amp}) for the energy and equation~(\ref{cc.eq}) for the expected cross correlation).

\begin{figure}
\centering
\includegraphics*[width=\linewidth]{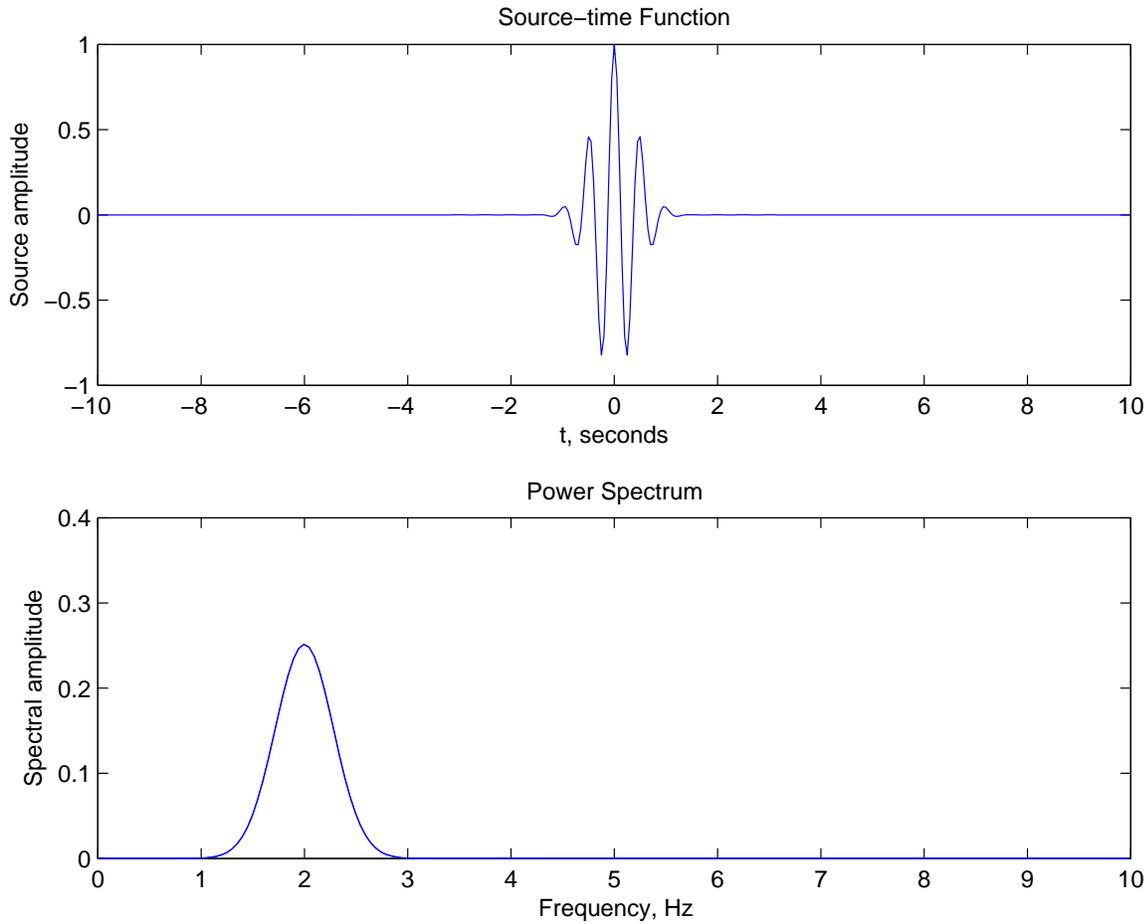}
\caption{Source-time function (upper panel) and the power spectrum ${\mathcal P}(\omega)$.\label{source}}
\end{figure}

\begin{figure}
\centering
\includegraphics*[width=\linewidth]{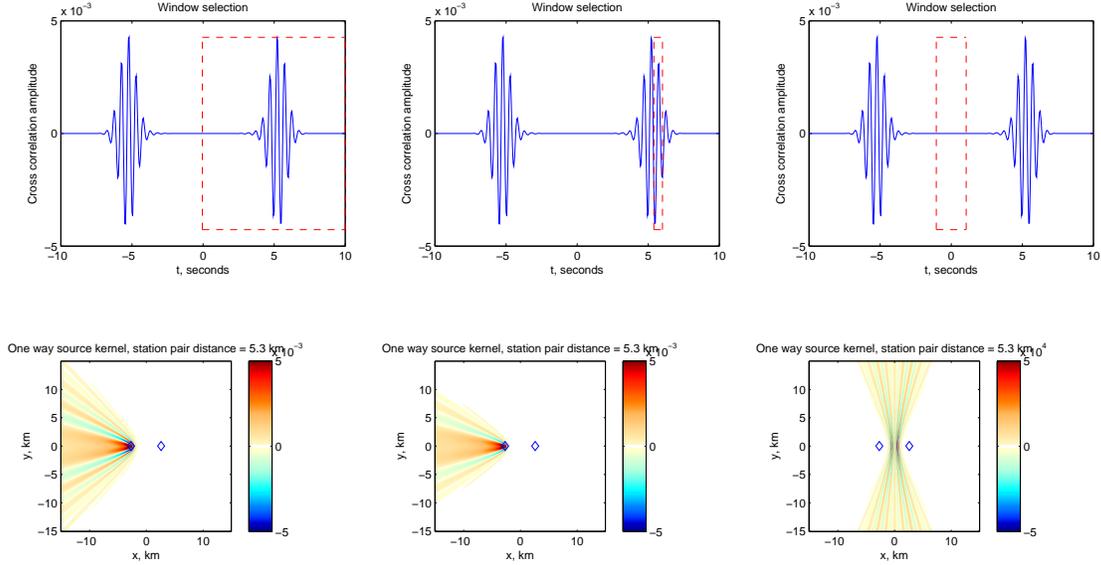}
\caption{Expected cross correlation (upper panels) and the attendant kernels (bottom panels).
Stations are marked by the diamond symbols. A wider measurement window implies that 
the the hyperbolae are thicker, as can be seen upon comparing lower left and middle panels. Note that when the window straddles the
zero time lag part of the cross correlation, the kernel will show sensitivity to sources along the bisecting line 
perpendicular to the path between the stations (bottom right). The part of the kernel along the line joining the two stations (i.e., $y=0$)
is sensitive to parts of the cross correlation corresponding to late times whereas the hyperbolae closest to the bisector (i.e., $x=0$)
are due to the zero-time-lag part of the cross correlation. The color scale has been saturated so as to enhance the visibility
of the relatively weak hyperbolae. \label{kernel.source}}
\end{figure}

We compute source-amplitude kernels (around a uniform distribution $\sigma = 1$)
in the temporal Fourier domain, using the exact functional form of Green's function~(\ref{green.hank}). 
The wavespeed is set to $c = 1$ km/s. The expected (limit) cross correlation contains symmetric positive and negative branches.
The power spectrum and its temporal representation are shown in Figure~\ref{source}. We use a temporal grid of 401 points
and the frequency spacing of 0.05 Hz, and hence a time window of 20 seconds, as in Figure~\ref{source}. The spacing
in the temporal grid is 0.05 seconds, implying a Nyquist frequency of 10 Hz. In order to compute the integral over frequency in equation~(\ref{kern.source}),
we precompute Hankel functions on a grid of $681\times681$ points resolving a square of size $[-40, 40]\times[-40, 40]$ km$^2$. 
These function values are then utilized to compute the expected cross correlation and kernels. For the examples in Figure~\ref{kernel.source}, we choose $\sigma(\bx) =1$, i.e., a
uniform distribution. 
In Figure~\ref{kernel.source}, we show examples of how the choice of the measurement affects the sensitivity kernel. On the upper panels,
the expected cross correlation for a point pair separated by a distance of 10.6 km. The dashed box indicates the choice of measurement, a
4 second window on the left panel and a 0.5 second window on the right. Kernels corresponding to these choices are shown immediately
below. A thicker hyperbola, indicative of a broader range of path differences, is seen on the kernel to the left (compared with the thinner hyperbolae on 
the right). Because we choose only the positive branch, the kernel shows sensitivity only to waves that first arrive at the station on the left 
and subsequently to the one on the right. Consequently, the hyperbolae point to the left.

{In general, the source-amplitude kernel depends indirectly on the choice of the initial spatial distribution of 
sources. Firstly, the cross correlation is obtained by evaluating an integral involving the spatial source distribution over space
 (Eq.~[\ref{cc.hank}]).
This predicted cross correlation is then used in the computation of the kernel (Eqs.~[\ref{kern.source}] and~[\ref{weight}]).
The function $W_{\alpha\beta}$, from equation~(\ref{weight}), assigns a frequency-dependent weight to the two Green's functions 
in the integral~(\ref{kern.source}). But since Green's functions are an inseparable mix of frequency and space (see Eq.~[\ref{green.hank}]), 
the kernel resulting from the evaluation of~(\ref{kern.source}) will possess a spatial dependence that reflects the source distribution. }

\begin{figure}
\centering
\includegraphics*[width=\linewidth]{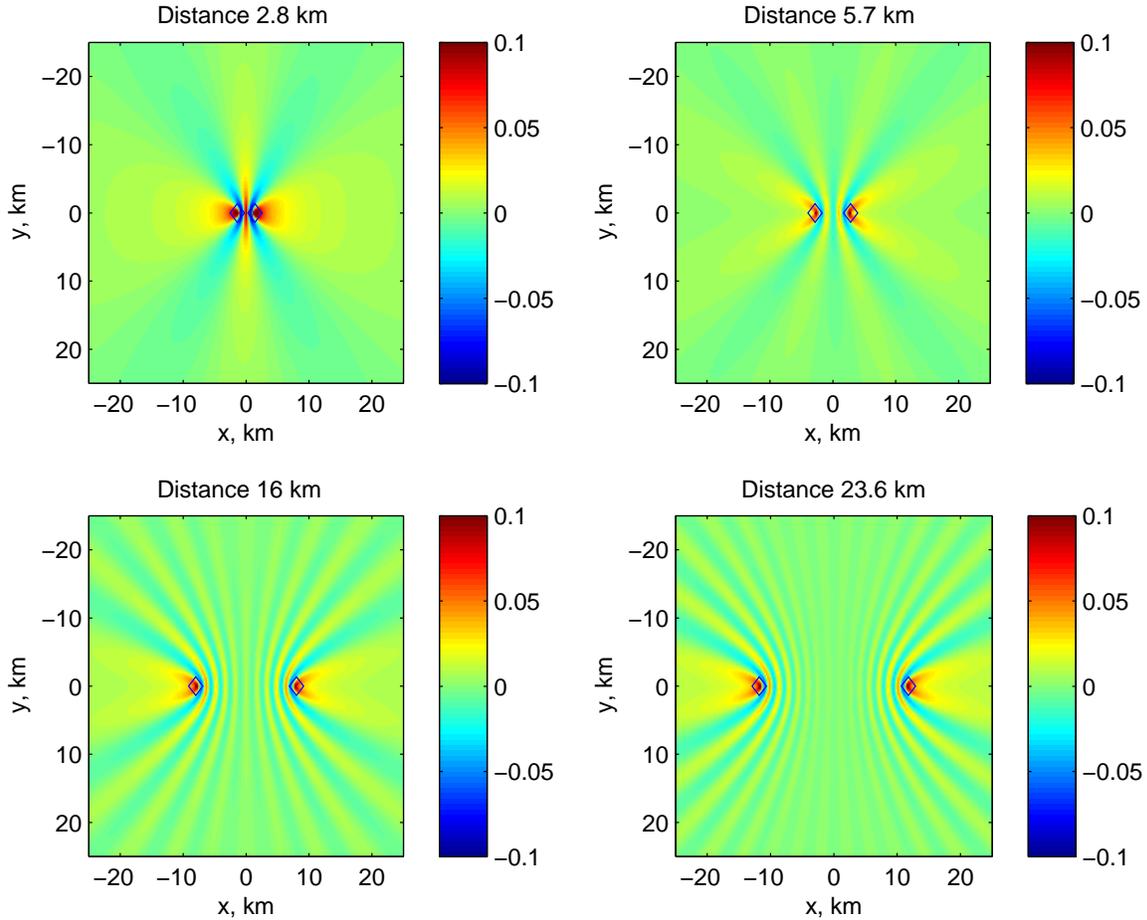}
\caption{{Source kernels as a function of interstation distance for a uniform source distribution. The measurement is the energy of the 
entire waveform. At small distances, the lateral size of the kernel is comparable
to the interstation distance whereas at very large distances, the sensitivity is restricted to a small range of azimuths around the
interstation path.}
\label{kerndist}}
\end{figure}

{Source-amplitude kernels as a function of interstation distance are graphed in Figure~\ref{kerndist}. It is seen
 that at small distances, the lateral size of the kernel is comparable
to the interstation distance whereas at very large distances, the sensitivity is restricted to a small range of azimuths around the
interstation path. There are disadvantages to using measurements at stations separated by small distances since they are only able to image
the source distribution in their vicinity, leading possibly to errors in the inversion. 
Further, if the distance between the stations is less than a wavelength, the cross correlation does not provide very much additional information
and should be removed from the analysis.
}

\begin{figure}
\centering
\includegraphics*[width=\linewidth]{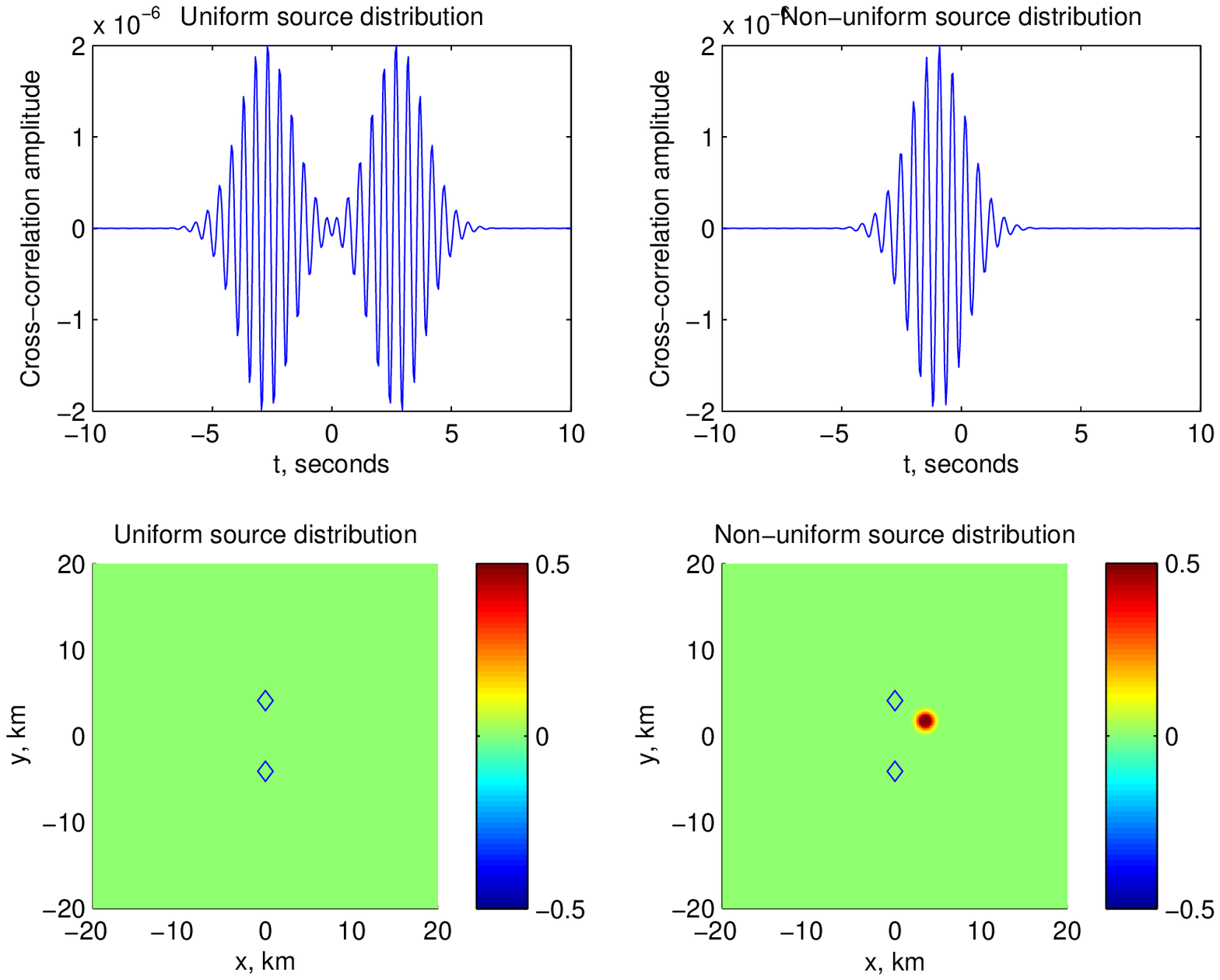}
\caption{{Source kernels as a function of frequency of the measurement. The source-time functions used
in computing the two sets of kernels are plotted on the bottom two panels.
The measurement is the energy of the 
entire waveform. When we use a more rapidly varying source-time function (right column), 
the kernel shows greater complexity and spatially sharper features. Much as in classical tomography, finer-scale images of 
 the source distribution may be obtained by using higher-frequency measurements.}
\label{kernfreq}}
\end{figure}

{In Figure~\ref{kernfreq}, we show kernels as a function of interstation distance and the temporal frequency of the measurement.
Higher-frequency waves lead to kernels with greater 
complexity and spatially sharper features. Much as in classical tomography, finer-scale images of 
 the source distribution may be obtained by using higher-frequency measurements. It is unclear how useful
 this will be when attempting to invert for oceanic microseisms since they occur at narrow frequency bands. In other words,
 if one were to incorporate higher or lower frequency measurements, away from the central microseism excitation frequency, would
 be of very limited utility in imaging. A more thorough study is needed in this regard.
}

\subsection{Non-uniform source distribution and the event kernel}\label{nonuni.dist}
\begin{figure}
\centering
\includegraphics*[width=\linewidth]{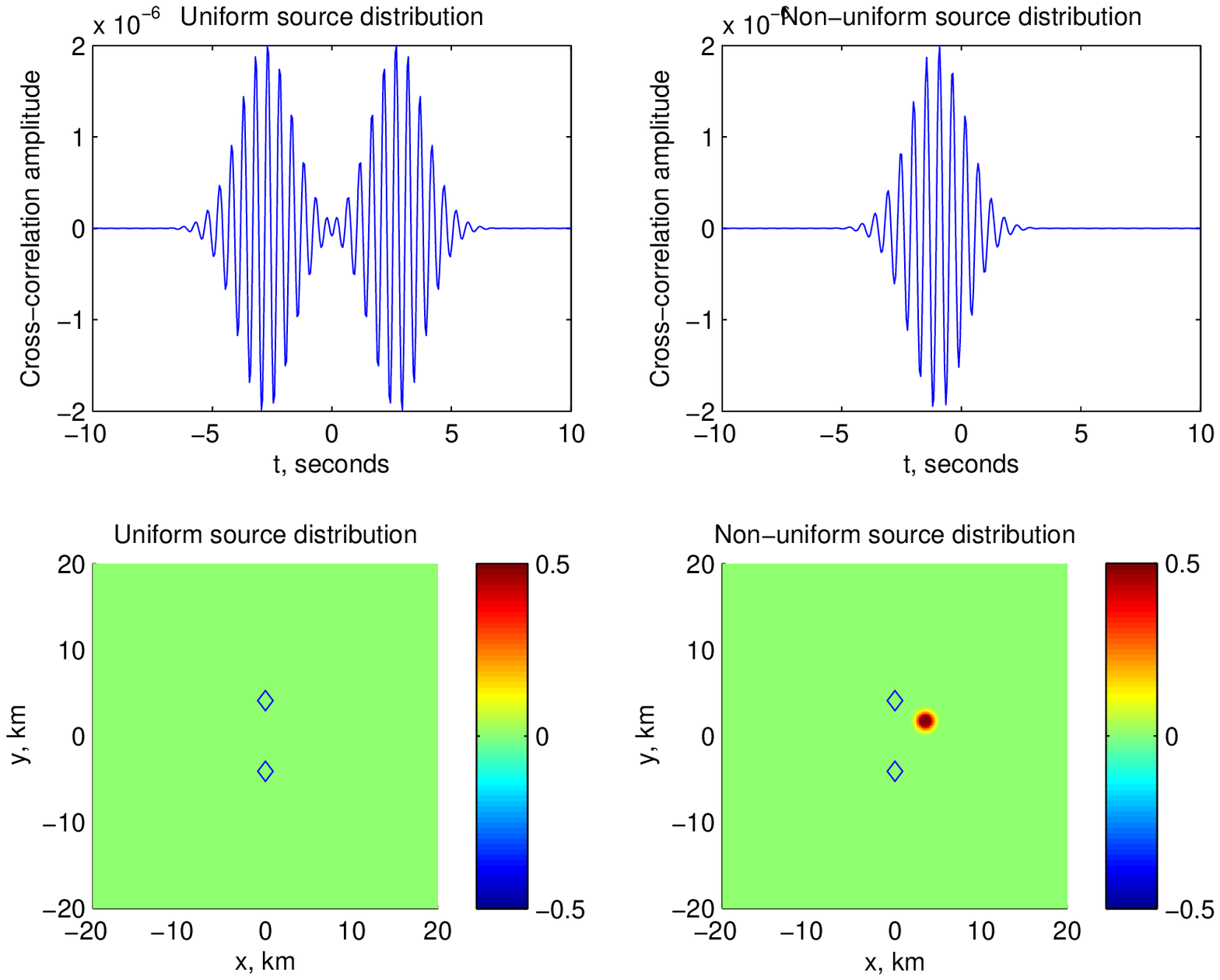}
\caption{Expected cross correlation (upper panels) and the source distributions (bottom panels).
Stations are marked by the diamond symbols. Pathological distributions (such as the lower right) of sources can cause large shifts
in the expectation value of the cross correlation because of the non-uniform manner in which waves illuminate the stations. 
In this case, {the only source being the spot}, is very close to the bisector line perpendicular 
to the path between the stations (i.e., $y=0$), implying that the path difference between the spot and the stations is nearly zero. The cross correlation is 
thus almost (but not quite) centered around zero time. 
\label{cc.comparison}}
\end{figure}

\begin{figure}
\centering
\includegraphics*[width=\linewidth]{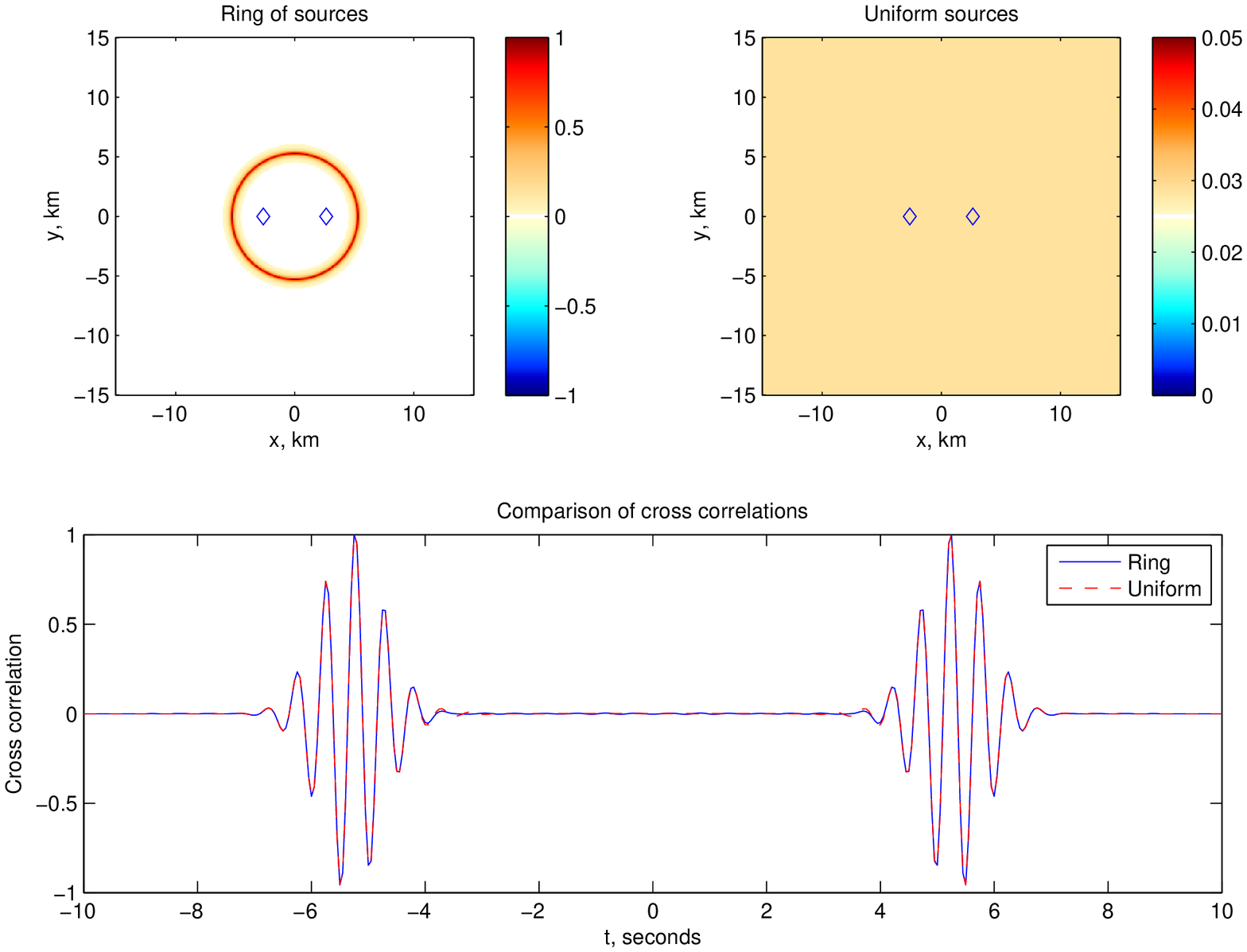}
\caption{The celebrated ``ring of sources surrounding a station pair" scenario (left panel) and uniform distribution of weaker amplitude (right panel). 
Cross correlations at the stations (symbols) due to these distributions are over plotted (bottom panel). {They are practically
indistinguishable, and is part of the reason why the ring configuration of sources has been studied so extensively.} \label{ring}}
\end{figure}

The presence of strong non-uniformities can render inaccurate results pertaining to the correspondence
between the cross correlation measurement and Green's functions along the station pair. The integral in~(\ref{cc.hank})
is over all space and provided the weight function $\sigma(\bx)$ is also uniform, the expectation value of the cross correlation
is well behaved, displaying features similar to classical tomographic arrivals, an instance of which is shown on the upper panels of Figure~\ref{cc.comparison}.
In fact, for the case considered here, the expected cross correlation can be shown to be a frequency modulation of Green's functions
along the path \citep[without wave attenuation; see][]{tromp10}.
However, when the source distribution $\sigma$ becomes more non-uniform, the expectation value of the cross correlation shifts away from
the elegant Green's function analog and adopts more complicated forms. A particularly stark example is when the sources lie along the bisector
line perpendicular to the path between the station-pair: the cross correlation in such a case will be centered around zero time, since the path
difference from the source to the stations is zero. We also consider a situation that has been studied extensively in past literature \citep[e.g.,][]{derode03, larose06, snieder10},
namely that of a ring of sources surrounding a station pair. Indeed, we find in Figure~\ref{ring} that the cross correlation owing to
a uniform distribution of sources is almost identical to that in the ring of sources scenario. As opposed to a discrete number of sources placed
at a certain radius around the station pair, we use a continuous annulus to represent the ring. The amplitude of the uniform distribution is substantially
smaller in order that the cross correlations from these two situations have the same energy.

\begin{figure}
\centering
\includegraphics*[width=\linewidth]{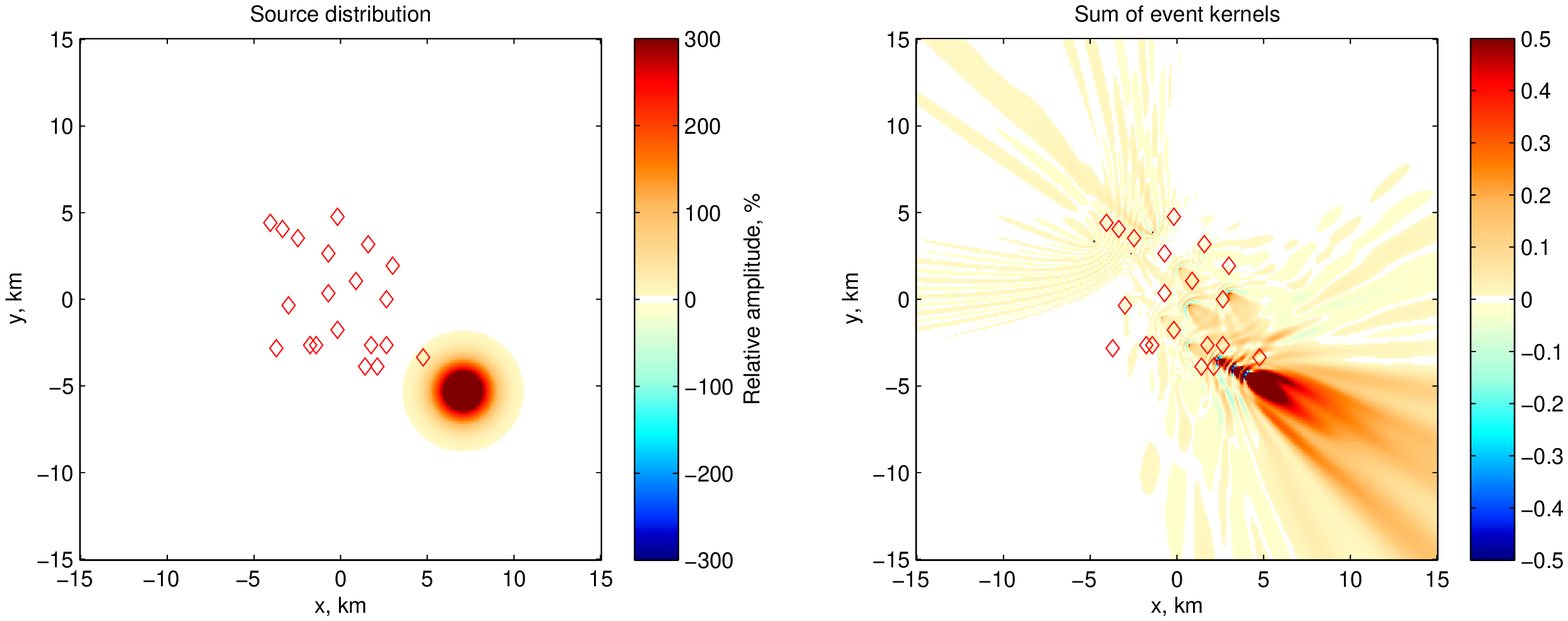}\vspace{-2cm}
\caption{Sum of event kernels. The left panel shows the ``{\it true}" source distribution, 
with respect to a nominal (uniform) value. Stations are marked by symbols. It contains a local spot of relatively large amplitude (500\% increase over the uniform value) on the
south-east quadrant. Because of the relative proximity of the station array to the sources, the event kernel
is able to roughly localize over the spot. {Since there is no coda (and complex structure) in this simple test case, the window 
encompasses the entire waveform.
Thus the energy of the entire cross correlation contributes to the construction of the image of the noise source distribution.}
\label{event}}
\end{figure}

In Figure~\ref{event}, we make the case for the imaging of anisotropic source distributions. The ``{\it true}" distribution is shown
on the left panel. The starting ``{\it synthetic}" model consists of a uniform distribution of sources whose amplitude is the same 
as that of the ``{\it true}" distribution away from the local spot of increased amplitude on the south-east quadrant. 
{We use the energy of the full envelope of the cross correlation, from zero to the classical surface-wave
arrival time.}
We compute the
misfit according to equation~(\ref{misf}). The event kernel associated with a particular station $\alpha$ is a sum of the point-to-point kernels
between that station and all other stations, i.e.,
\begin{equation}
K_\alpha(\bx) = \sum_\beta \ln\frac{A^{\rm obs}_{\alpha\beta}}{A^{\rm syn}_{\alpha\beta}}\,K_{\alpha\beta}(\bx),
\end{equation}
where $A^{\rm obs}$ signifies the ``{\it true}" energies. The sum of all the event kernels associated with a network of stations provides
a way to update the model, and is $\sum_\alpha K_{\alpha}(\bx)$. As discussed in \cite{tromp10}, the cost of the inversion 
scales with the number of ``master" pixels $\alpha$ and is independent of the number of ``slaves" $\beta$. However because
we use a translationally invariant background model, the computation of kernels is cheap and so we may include as many master pixels
as we desire. In this case, we stop at 20 stations, i.e., 20 master and 19 slave pixels. 

\begin{figure}
\centering
\includegraphics*[width=\linewidth]{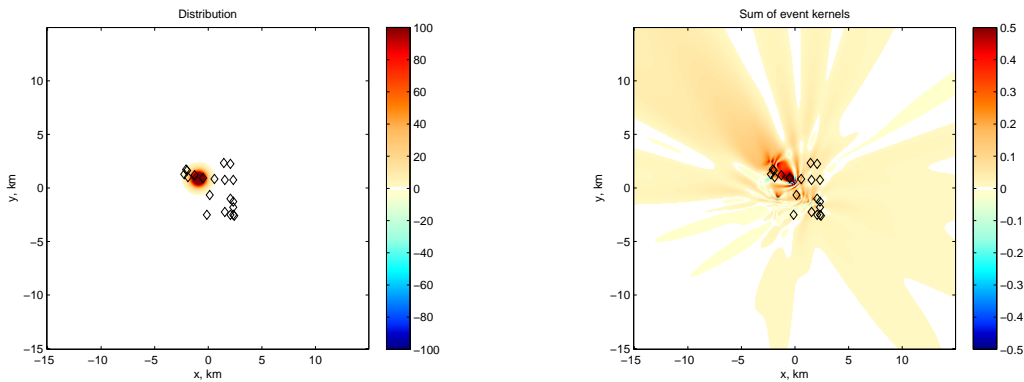}\vspace{-1cm}
\caption{Sum of event kernels. The left panel shows the ``{\it true}" source distribution, 
with respect to a nominal (uniform) value. Stations are marked by symbols. We include a local spot of relatively large amplitude (200\% increase over the uniform value) 
amid the station array. Because stations surround the anomaly, they are able to accurately locate the distribution. {Since there is no coda (and complex structure) in this simple test case, the window 
encompasses the entire waveform.
Thus the energy of the entire cross correlation contributes to the construction of the image of the noise source distribution.}
\label{eventcent}}
\end{figure}

\begin{figure}
\centering
\includegraphics*[width=\linewidth]{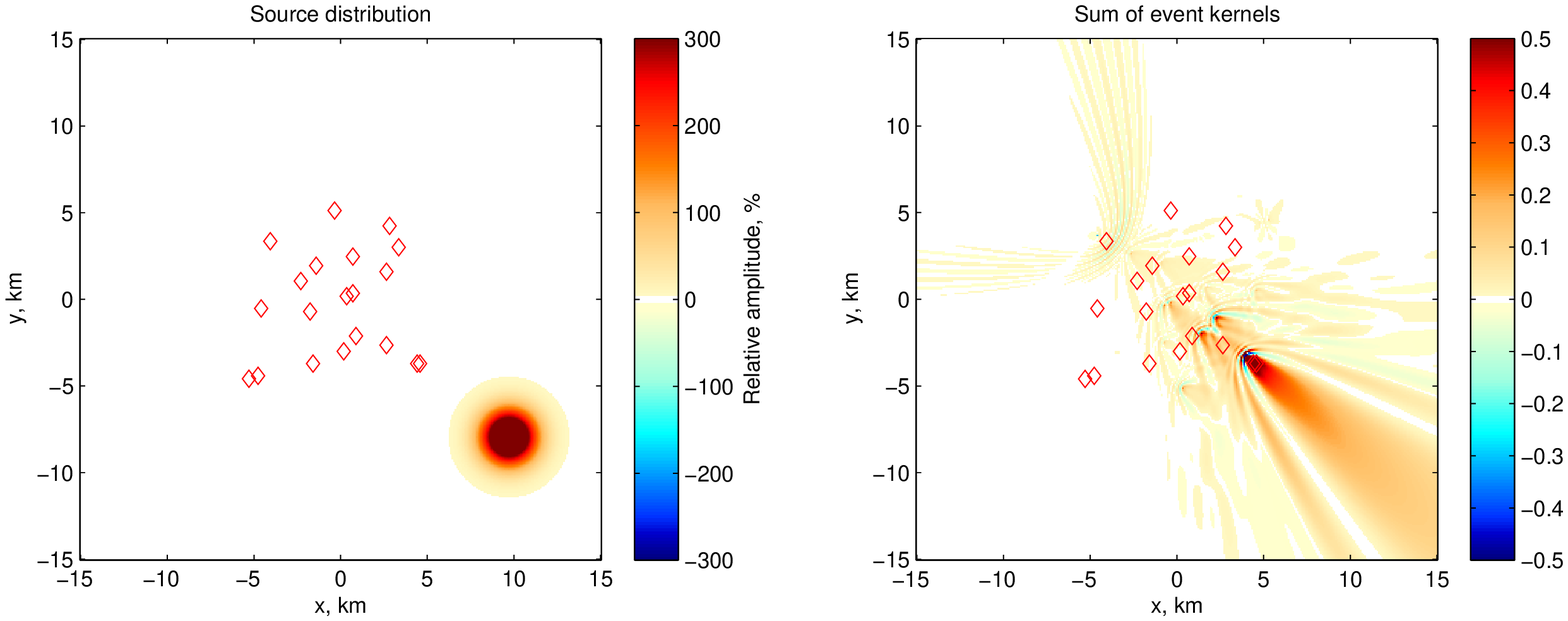}\vspace{-2cm}
\caption{Sum of event kernels. The left panel shows the ``{\it true}" source distribution, 
with respect to a nominal (uniform) value. Stations are marked by symbols. We include a local spot of relatively large amplitude (500\% increase over the uniform value) moved
farther away from the station array than in Figure~\ref{event}. Evidently, the stations are too far away from the sources to accurately locate the distribution.  {Since there is no coda (and complex structure) in this simple test case, the window 
encompasses the entire waveform.
Thus the energy of the entire cross correlation contributes to the construction of the image of the noise source distribution.}
\label{event2}}
\end{figure}

In Figure~\ref{event}, we show the sum of event kernels corresponding to this set of ``data" and ``synthetics". The kernels neatly focus onto
the area where the source amplitude is locally large (by 500\% in comparison to the value away from this spot). In order to image this localized spot better,
we would require a greater coverage by the array, i.e., an array that surrounds the spot, as shown in Figure~\ref{eventcent}. Note that the farther away the anomalous source activity is from the array, 
thus diminishes the ability to discern their location. An example of such a situation is shown in Figure~\ref{event2}. One may interpret it as the sources being far enough away
that when the waves arrive at the stations, their curvature ($\propto \Delta_S^{-1}$, $\Delta_S$ being the distance from the source) is so small that they appear as 
plane waves, and information about the source location is thus lost. In order to perceive such small curvature, a network of stations placed far apart ($\propto \Delta_S$) 
becomes necessary. {An additional reason for the spatially localized sensitivity is that the source kernel is of greatest amplitude
along the interstation paths and in the vicinities of the stations.
Conceptually, imaging of the noise source distribution is not very different from that of inverting for an earthquake source; both require
appropriate choices for measurements and a good network of stations. 
Subsequently, by studying the inverted source distribution, we may arrive at the conclusion that the distribution is too far away
to image. In this case the procedure is essentially no different from beamforming \citep[e.g.,][]{stehly06}. 
However, if there were to be more information in the wavefield (by using different time windows
and some intrinsic curvature properties of the waves), then this method will be able to utilize it to produce better
quality images of the noise source distribution than beamforming.}

{
Rapid temporal variations in the source distribution may also be lost when the cross correlations are averaged over long times. However, this is independent of the technique used, i.e., beamforming or an adjoint method (used here). 
Therefore it is useful to apply the sorts of methods described here since it maximally utilizes wavefield information. }

\section{Scattering}\label{scatter}
{In the framework of correlation tomography, scattering kernels can be substantially more complicated. They are also
intrinsically different in flavor from classical banana-doughnut kernels, containing additional hyperbolic features which
represent sensitivity to sources at disparate spatial locations. Much as in Figures~\ref{cc_general_sources} and~\ref{cc_here}, 
we attempt here to graphically explain the physics of scattering kernels for noise measurements.}

\begin{figure}
\centering
\includegraphics*[width=\linewidth]{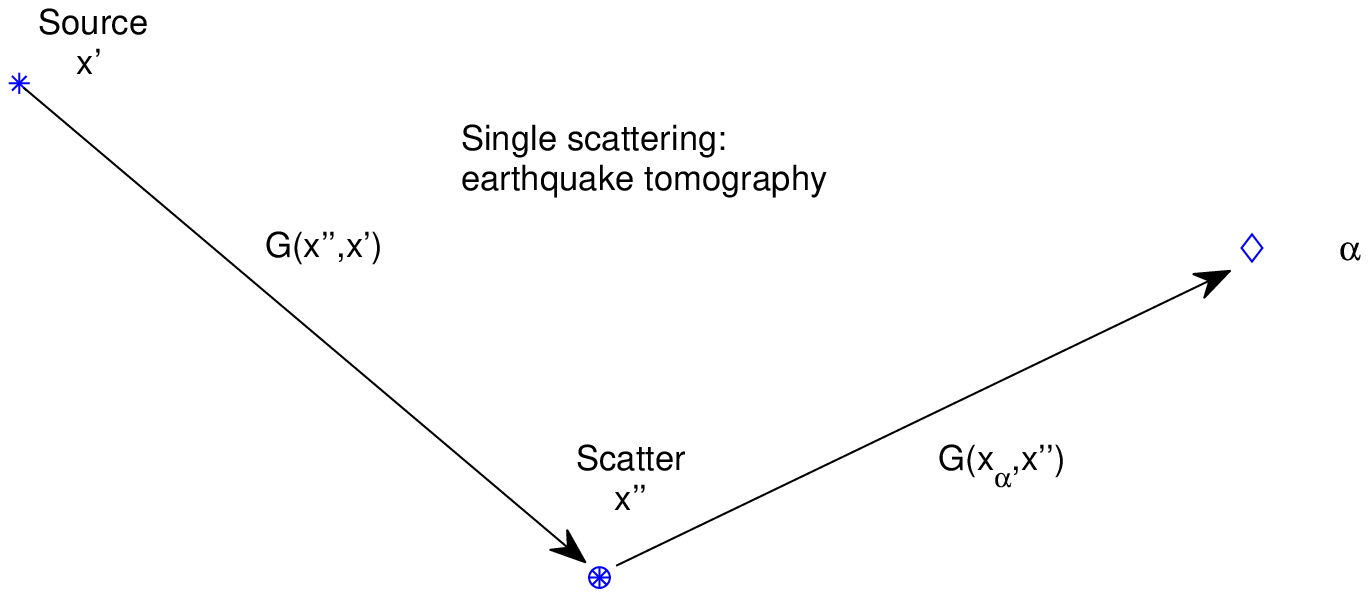}
\caption{Single scattering in the classical earthquake tomography case. A 
source at ${\bx'}$ excites a wave that propagates to $\bx{''}$ where it scatters, acts as a source, propagating finally to station $\alpha$. It is substantially
simpler than the cross correlation measurement, which is depicted in Figures~\ref{scattering1} and~\ref{scattering2}. Only
two Green's functions are required to model single scattering in this scenario. \label{earthquake}}
\end{figure}

Variations to the limit cross correlation~(\ref{expect}) are given by
\begin{equation}
\langle\delta{\mathcal C}_{\alpha\beta}\rangle = \langle \phi^*(\bx_\alpha)\,\delta\phi(\bx_\beta)\rangle + \langle \delta\phi^*(\bx_\alpha)\,\phi(\bx_\beta)\rangle + O(\delta\phi^2),\label{varcc}
\end{equation}
in which, keeping with convention, we do not explicitly state the dependence on frequency $\omega$. Hitherto, we have ignored scattering terms but in 
this section, we describe their mathematical structure. For a given wave operator ${\mathcal L}$ and the corresponding wavefield $\phi$ satisfying
\begin{equation}
{\mathcal L}\phi = S,
\end{equation}
the first Born approximation \citep[e.g.,][]{hudson77,wu85} describing the singly scattered wavefield $\delta\phi$ owing to perturbations to the operator, $\delta{\mathcal L}$, is 
\begin{equation}
\delta({\mathcal L}\phi) = {\mathcal L}\delta\phi + \delta{\mathcal L}\phi = 0,
\end{equation}
where we assume that the source distribution is known exactly. We have 
\begin{equation}
{\mathcal L}\delta\phi = - \delta{\mathcal L}\phi,
\end{equation}
which upon using Green's theorem (Eq.~[\ref{green.eq}]) for the wavefield, we obtain
\begin{equation}
\delta\phi(\bx) = -\int d\bx{''} G(\bx,\bx{''}) [\delta{\mathcal L}\phi],
\end{equation}
where $\phi = \phi(\bx{''})$ and which may be rewritten in terms of the source distribution $S(\bx')$ as
\begin{equation}
\delta\phi(\bx) = -\int d\bx{''} G(\bx,\bx{''})\, \delta{\mathcal L}\left(\int d\bx'\,G(\bx{''},\bx')\,S(\bx')\right).
\end{equation}
This equation states that a source creates a wave at $\bx'$, which propagates to a point $\bx{''}$ as described by
Green's function along that path, is singly scattered according to $\delta{\mathcal L}$ and subsequently acts as a source, eventually propagating to measurement point
$\bx$. This is the framework in which classical tomographic scattering is studied (shown in Figure~\ref{earthquake}). Substituting this into equation~(\ref{varcc}),
\begin{eqnarray}
&&\langle\delta{\mathcal C}_{\alpha\beta}\rangle = -\langle \left[\int d\bx \,G^*(\bx_\alpha,\bx)\,S^*(\bx)\right] \left\{\int d\bx{''} G(\bx_\beta,\bx{''})\, \delta{\mathcal L}\left(\int d\bx'\,G(\bx{''},\bx')\,S(\bx')\right)\right\}\rangle\label{varcc.1}\\
&&\mbox{} -\langle \left\{\int d\bx{''} G^*(\bx_\alpha,\bx{''})\, \delta{\mathcal L}^*\left(\int d\bx'\,G^*(\bx{''},\bx')\,S^*(\bx')\right)\right\}\left[\int d\bx \,G(\bx_\beta,\bx)\,S(\bx)\right]\rangle.\nonumber
\end{eqnarray}

\begin{figure}
\centering
\includegraphics*[width=\linewidth]{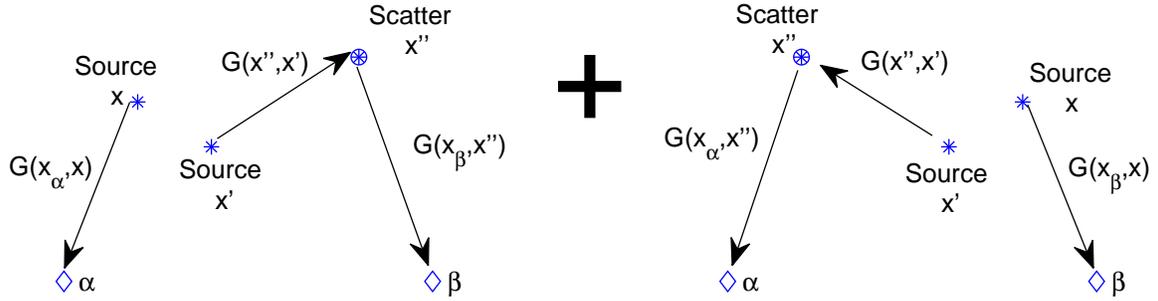}
\caption{Scattering as captured by cross correlations for a general source distribution (similar to Figure~\ref{cc_general_sources}) with non-zero spatial covariance. The first contribution (first line of Eq.~[\ref{varcc.1}])
consists of a correlation between a wave generated at point $\bx$, propagated to $\alpha$ and a wave generated at $\bx'$, scattered according to
perturbation $\delta{\mathcal L}$ at point $\bx{''}$, propagated to measurement point $\beta$. The second contribution (second line of Eq.~[\ref{varcc.1}])
is essentially the same except with points $\alpha, \beta$ reversed. This is the reason why three Green's functions are needed to formally
interpret the measurement in terms of single-scattering theory.\label{scattering1}}
\end{figure}

The terms in the flower brackets denote the scattering contributions and the terms within the square brackets show the direct wave arrival from the source to the observation points.
Because we truncate the Born approximation to one term, i.e., considering only contributions from single scattering processes, the variation of the cross correlation consists
of a direct wave propagating from a source to one of the stations, correlated with a singly scattered wave that propagates to the other measurement point. Akin to \citet{gizon02}, 
we diagrammatically show the formal interpretation of the measurement in Figure~\ref{scattering1}. In contrast to single-scattering theory applied to classical tomographic
wavefield measurements (shown in Figure~\ref{earthquake}), in which two Green's functions appear, the higher-order correlation measurement (regardless of whether these are ``noise" or earthquake sources) requires the evaluation of three Green's functions.

\begin{figure}
\centering
\includegraphics*[width=\linewidth,clip=]{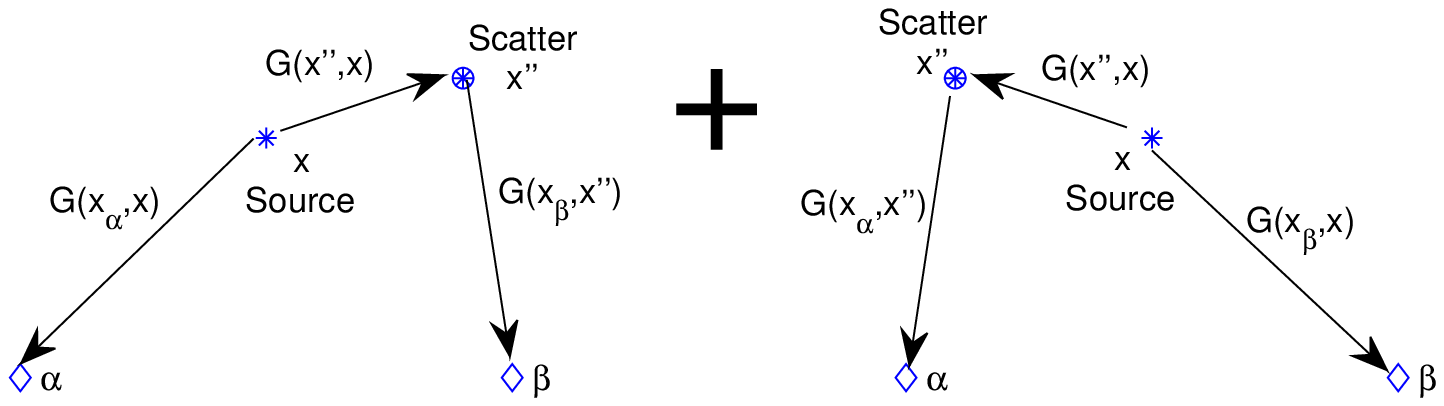}
\caption{Scattering as captured by cross correlations for a source distribution with zero spatial covariance (similar to Figure~\ref{cc_here}) . The first contribution (first line of Eq.~[\ref{varcc.2}])
consists of a correlation between a wave generated at point $\bx$ propagated to points $\alpha$ and $\bx{''}$, with the latter scattered according to
perturbation $\delta{\mathcal L}$ at point $\bx{''}$ and eventually propagated to measurement point $\beta$. The second contribution (second line of Eq.~[\ref{varcc.2}])
is essentially the same except with points $\alpha, \beta$ reversed. 
This is the reason why three Green's functions are needed to formally
interpret the measurement in terms of single-scattering theory. The only difference between this figure and Figure~\ref{scattering1} is that the
contributions to the cross correlation are from the same source point (i.e., when $\bx = \bx'$). \label{scattering2}}
\end{figure}

Upon invoking the assumption of spatially uncorrelated sources, i.e., $\langle S(\bx)\,S^*(\bx')\rangle  = \delta(\bx - \bx')\, \sigma(\bx)\,{\mathcal P}(\omega)$, we obtain
\begin{eqnarray}
&&\langle\delta{\mathcal C}_{\alpha\beta}\rangle = -\int d\bx{''} G(\bx_\beta,\bx{''})\, \delta{\mathcal L}\left(\int d\bx'\,G(\bx{''},\bx)\,G^*(\bx_\alpha,\bx)\,\sigma(\bx)\,{\mathcal P}\right)\label{varcc.2}\\
&&\mbox{} -\int d\bx{''} G^*(\bx_\alpha,\bx{''})\, \delta{\mathcal L}^*\left(\int d\bx\,G^*(\bx{''},\bx)\,G(\bx_\beta,\bx)\,\sigma(\bx)\,{\mathcal P}\right),\nonumber
\end{eqnarray}
which produces a scattering diagram similar to Figure~\ref{scattering1}, except with coinciding points $\bx, \bx'$, as shown in Figure~\ref{scattering2}. {These kernels are indeed more difficult to compute than in the classical tomography case, and evidently
require the evaluation of three Green's functions. The physics of these kernels is also conceptually different from the 
classical case.}

\section{The sensitivity of cross-correlation energies to attenuation}\label{attenuation}
The topic of imaging wave attenuation using cross-correlation energies is a topic of interest \citep[e.g.,][]{cupillard10, weaver11, prieto11, tsai11}. 
The challenge is to accurately interpret enhanced decrements in cross-correlation energies amid effects of geometrical spreading with distance and
wave-speed heterogeneities. Evidently, the distribution of sources significantly
influences the conclusion of any inverse problem, and the problem of the determination of wave attenuation is no different.
{This suggests that the strategy typically followed in
earthquake tomography, which is to first invert for the source and subsequently for structure (perhaps iteratively), may be applied equally
to noise measurements. The standard trade-off between source and structure affects the interpretation of noise measurements as well.}

\begin{figure}
\centering
\includegraphics*[width=\linewidth,clip=]{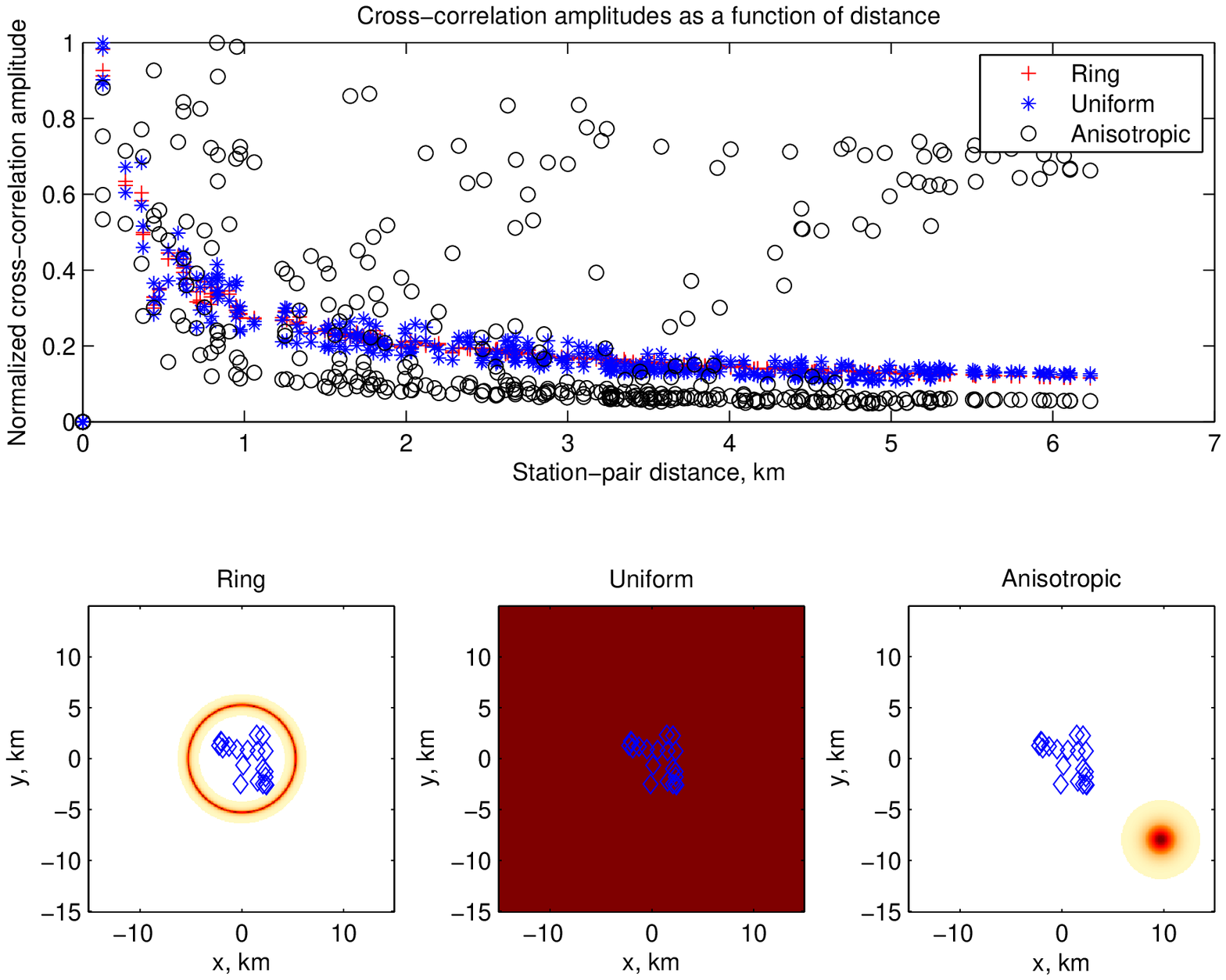}
\caption{Cross-correlation energy (or amplitude; Eq.~[\ref{syn.amp}]) as a function of distance between station pairs (the network shown in the lower panels). Stations, marked by
diamonds, are illuminated by a ring of sources (left), uniform sources (middle) and by a highly anisotropic distribution (right).
The scatter in energies is entirely due to geometric spreading and source distribution anisotropy (for the anisotropic case).
Amplitudes of every branch of every cross correlation are plotted (independent of orientation) in the upper panel for the three cases. 
They are normalized in all three cases such that the greatest value is 1. 
Determining the source distribution prior to interpretation is strongly tied to the accurate interpretation of these measurements. 
{However, it may be noted that in the anisotropic case, the normalization could be chosen to
be azimuthally dependent, resulting in energy 	shifts better suited to fitting \citep[][]{prieto09}.}  \label{ampvar}}
\end{figure}

In Figure~\ref{ampvar}, we show the variation in energy (defined as the energy of the cross-correlation branch~(\ref{syn.amp}), positive
or negative) of the cross correlation as a function of distance between the station pair for three source distributions. The variation in
energy is entirely due to geometrical spreading and source distribution anisotropies. Note that the background model has no attenuation in this case. {The scatter in cross correlation energies for the anisotropic case may possibly be reduced by choosing
an azimuthally varying normalization \citep{prieto09}.}

{The ring of sources case is seen to be different from the uniform distribution. This is because the in an attenuating medium, 
sources from farther away contribute less to the cross correlation. Therefore, the cross correlation energies in the
ring and uniform cases converge to different expectation values, whose difference increases with the extent of attenuation.
 }

\begin{figure}
\centering
\includegraphics*[width=\linewidth,clip=]{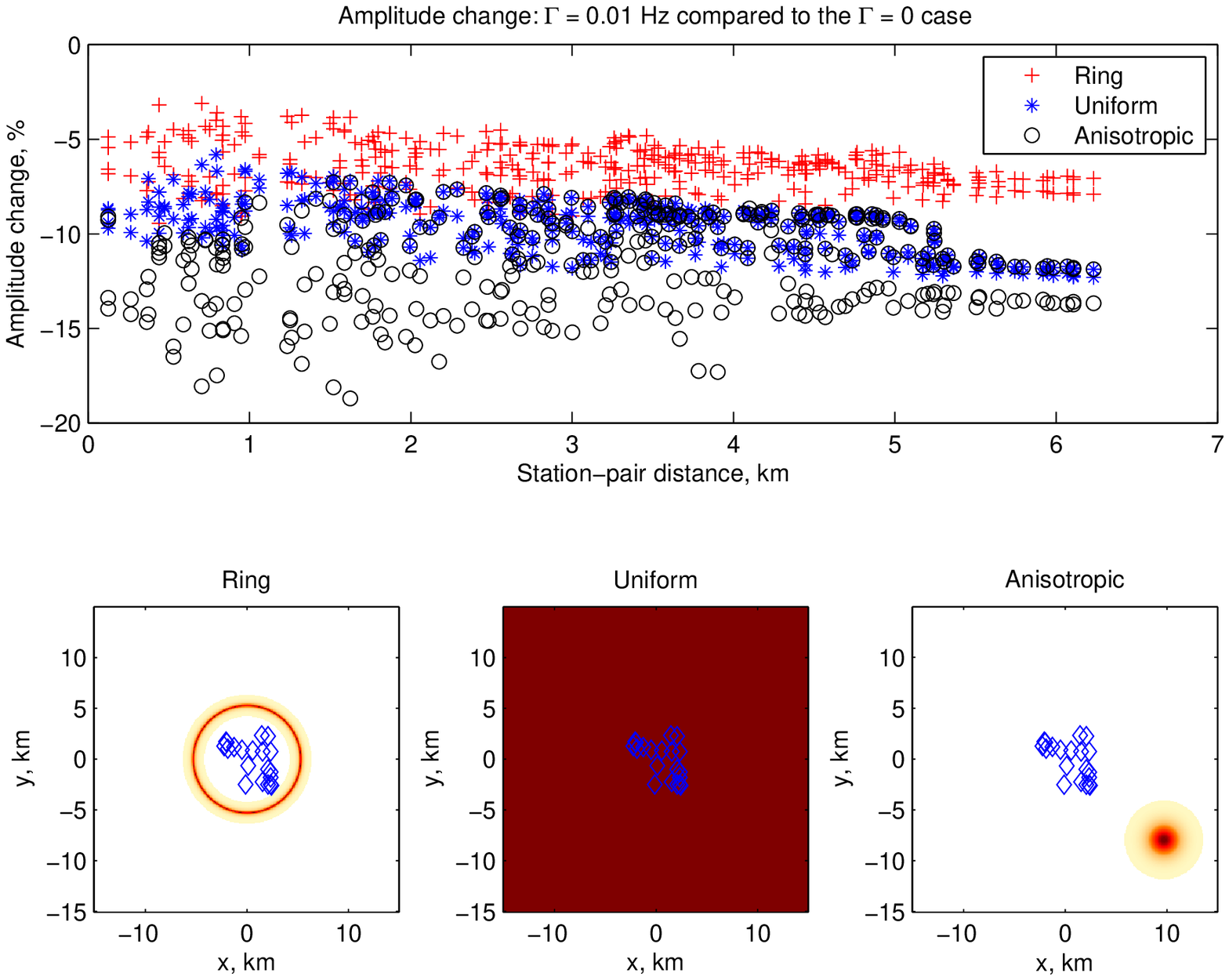}
\caption{Change in energies (amplitudes) due to the introduction of wave attenuation of cross correlations for the network and source distributions of Figure~\ref{ampvar}.  
We use a damping rate of $\Gamma = 0.01$ Hz. We plot the relative energy change of every branch of every cross correlation as a function of station-pair distance. 
The anisotropic distribution shows the largest changes in energies - and this is because upon introducing attenuation, waves which travel from the patch of sources arrive with diminished energies by the time they reach the network. 
The effect is less severe in the ring of sources case, where the ring
is close to the network. 
In any case, the knowledge of the source distribution is critical to inferring properties of the underlying medium. 
In the scenario of a uniform distribution of sources (plus symbols), it certainly appears that the energy reduction due to attenuation is a robust 
effect and may be used in a straightforward manner to infer wave attenuation. \label{ampvardelta}}
\end{figure}

We also characterize the significance of a finite quality factor on energies when the network is illuminated by
the source distributions of Figure~\ref{ampvar}. We use a damping rate of 0.01 Hz, or a quality factor of roughly 150. Wave attenuation is modeled
via solutions of the damped simple harmonic oscillator, i.e., operator~(\ref{2deq}) with a damping term
\begin{equation}
\partial_t^2 \phi + \Gamma\,\partial_t \phi - c^2\nabla^2\phi = 0,
\end{equation}
where attenuation $\Gamma$ has units of Hertz. Green's function for this operator is then
\begin{equation}
G(\bx,\bx') = \frac{i}{4} H^{(1)}\left(\frac{p\,\omega}{c}|\bx-\bx'|\right),
\end{equation}
where the factor $p = \sqrt{ 1 +i\Gamma/\omega}$. Amplitudes evidently change but in entirely different ways, depending on the source distribution, as displayed in Figure~\ref{ampvardelta}. We show
the percentage change in cross correlation energies due to the introduction of a spatially constant wave attenuation of 0.01 Hz. Figure~\ref{ampvardelta}
demonstrates that wave energies are indeed sensitive to attenuation, but extracting this information is subject to accurate knowledge of the source distribution.


\section{Conclusions}\label{conclude}
Cross correlations are intrinsically more complex than classically used wavefield displacements.
There are fundamental and meaningful differences between these measurements, which have
consequences for the eventual solution of inverse problems.
Because seismology is a precision science, it is important to formally interpret these measurements
and capture their essence as fully as possible. 
Using a simple 2-D example, we have endeavored to delve into the physics of the cross correlation measurement. 
A goal of this article was to demonstrate the utility and ease of studying distributions of sources and posing the problem in terms of the expectation value of
the relevant measurable. We make a case for the imaging of source distributions using measurements of cross-correlation energies.
The dependence of these
energies on station-pair distance and on wave attenuation is also touched upon. The influence of the source distribution on the 
energy measurement is demonstrated; cross-correlation energies unquestionably contain information about wave attenuation (primarily 
within the network) but it is hard to interpret them accurately without knowledge of the sources.

\section*{Acknowledgements}
S. M. H. is funded by NASA grant NNX11AB63G. S. M. H. thanks P. Cupillard, L. Stehly, R. Modrak and
two anonymous referees for considered comments and encouragement.

\bibliographystyle{gji}

\bibliography{noise}

\begin{thebibliography}{40}
\expandafter\ifx\csname natexlab\endcsname\relax\def\natexlab#1{#1}\fi

\bibitem[{Aki} \& {Richards}(1980)]{aki80}
{Aki}, K. \& {Richards}, P.~G., 1980.
\newblock {\it {Quantitative Seismology, Theory and Methods}\/}, {W. H.
  Freeman, San Francisco, California, USA}.

\bibitem[{Brenguier} et~al.(2007){Brenguier}, {Shapiro}, {Campillo},
  {Nercessian}, \& {Ferrazzini}]{campillo07}
{Brenguier}, F., {Shapiro}, N.~M., {Campillo}, M., {Nercessian}, A., \&
  {Ferrazzini}, V., 2007.
\newblock {3-D surface wave tomography of the Piton de la Fournaise volcano
  using seismic noise correlations}, {\it {Geophys. Res. Lett.}\/}, {\bf 34},
  2305.

\bibitem[Brenguier et~al.(2008)Brenguier, Campillo, Hadziioannou, Shapiro,
  Nadeau, \& Larose]{BrenguierSCI08}
Brenguier, F., Campillo, M., Hadziioannou, C., Shapiro, N., Nadeau, R., \&
  Larose, E., 2008.
\newblock Postseismic relaxation along the san andreas fault at parkfield from
  continuous seismological observations, {\it Science\/}, {\bf 321}(5895),
  1478--1481.

\bibitem[Chevrot et~al.(2007)Chevrot, Sylvander, Benhamed, Ponsolles,
  Lef{\`e}vre, \& Paradis]{ChevrotJGR07}
Chevrot, S., Sylvander, M., Benhamed, S., Ponsolles, C., Lef{\`e}vre, J., \&
  Paradis, D., 2007.
\newblock Source locations of secondary microseisms in western europe: Evidence
  for both coastal and pelagic sources, {\it J. Geophys. Res.\/}, {\bf 112},
  B11301.

\bibitem[{Cupillard} \& {Capdeville}(2010)]{cupillard10}
{Cupillard}, P. \& {Capdeville}, Y., 2010.
\newblock {On the amplitude of surface waves obtained by noise correlation and
  the capability to recover the attenuation: a numerical approach}, {\it
  Geophysical Journal International\/}, {\bf 181}, 1687--1700.

\bibitem[{Cupillard} et~al.(2011){Cupillard}, {Stehly}, \&
  {Romanowicz}]{cupillard11}
{Cupillard}, P., {Stehly}, L., \& {Romanowicz}, B., 2011.
\newblock {The one-bit noise correlation: a theory based on the concepts of
  coherent and incoherent noise}, {\it Geophysical Journal International\/},
  {\bf 184}, 1397--1414.

\bibitem[{Dahlen} \& {Baig}(2002)]{dahlen02}
{Dahlen}, F.~A. \& {Baig}, A.~M., 2002.
\newblock {Fr{\'e}chet kernels for body-wave amplitudes}, {\it Geophysical
  Journal International\/}, {\bf 150}, 440--466.

\bibitem[{Derode} et~al.(2003){Derode}, {Larose}, {Tanter}, {de Rosny},
  {Tourin}, {Campillo}, \& {Fink}]{derode03}
{Derode}, A., {Larose}, E., {Tanter}, M., {de Rosny}, J., {Tourin}, A.,
  {Campillo}, M., \& {Fink}, M., 2003.
\newblock {Recovering the Green's function from field-field correlations in an
  open scattering medium (L)}, {\it Acoustical Society of America Journal\/},
  {\bf 113}, 2973--2976.

\bibitem[{Duvall} et~al.(1993){Duvall}, {Jefferies}, {Harvey}, \&
  {Pomerantz}]{duvall}
{Duvall}, Jr., T.~L., {Jefferies}, S.~M., {Harvey}, J.~W., \& {Pomerantz},
  M.~A., 1993.
\newblock {Time-distance helioseismology}, {\it \nat\/}, {\bf 362}, 430--432.

\bibitem[{Fichtner} et~al.(2006){Fichtner}, {Bunge}, \& {Igel}]{fichtner06}
{Fichtner}, A., {Bunge}, H.-P., \& {Igel}, H., 2006.
\newblock {The adjoint method in seismology}, {\it Physics of the Earth and
  Planetary Interiors\/}, {\bf 157}, 86--104.

\bibitem[{Fleury} et~al.(2010){Fleury}, {Snieder}, \& {Larner}]{snieder10}
{Fleury}, C., {Snieder}, R., \& {Larner}, K., 2010.
\newblock {General representation theorem for perturbed media and application
  to Green's function retrieval for scattering problems}, {\it Geophysical
  Journal International\/}, {\bf 183}, 1648--1662.

\bibitem[Froment et~al.(2010)Froment, Campillo, Roux, Gou{\'e}dard, Verdel, \&
  Weaver]{FromentGEO10}
Froment, B., Campillo, M., Roux, P., Gou{\'e}dard, P., Verdel, A., \& Weaver,
  R., 2010.
\newblock Estimation of the effect of nonisotropically distributed energy on
  the apparent arrival time in correlations, {\it Geophysics\/}, {\bf 75},
  SA85.

\bibitem[{Gizon} \& {Birch}(2002)]{gizon02}
{Gizon}, L. \& {Birch}, A.~C., 2002.
\newblock {Time-Distance Helioseismology: The Forward Problem for Random
  Distributed Sources}, {\it \apj\/}, {\bf 571}, 966--986.

\bibitem[{Hanasoge} et~al.(2011){Hanasoge}, {Birch}, {Gizon}, \&
  {Tromp}]{hanasoge11}
{Hanasoge}, S.~M., {Birch}, A., {Gizon}, L., \& {Tromp}, J., 2011.
\newblock {The Adjoint Method Applied to Time-distance Helioseismology}, {\it
  \apj\/}, {\bf 738}, 100.

\bibitem[{Hudson}(1977)]{hudson77}
{Hudson}, J.~A., 1977.
\newblock {On the equations of elastodynamics}, {\it Geophysical Journal
  International\/}, {\bf 48}, 521--524.

\bibitem[{Kedar} \& {Webb}(2005)]{kedar05}
{Kedar}, S. \& {Webb}, F.~H., 2005.
\newblock {The ocean's seismic hum}, {\it {Science}\/}, {\bf 307}, 682--683.

\bibitem[{Larose} et~al.(2006){Larose}, {Montaldo}, {Derode}, \&
  {Campillo}]{larose06}
{Larose}, E., {Montaldo}, G., {Derode}, A., \& {Campillo}, M., 2006.
\newblock {Passive imaging of localized reflectors and interfaces in open
  media}, {\it Applied Physics Letters\/}, {\bf 88}(10), 104103.

\bibitem[{Larose} et~al.(2008){Larose}, {Derode}, {Roux}, \&
  {Campillo}]{larose08}
{Larose}, E., {Derode}, A., {Roux}, P., \& {Campillo}, M., 2008.
\newblock {Convergence of correlations in multiply scattering media}, {\it
  Acoustical Society of America Journal\/}, {\bf 123}, 3931.

\bibitem[{Longuet-Higgins}(1950)]{longuet50}
{Longuet-Higgins}, M.~S., 1950.
\newblock {A Theory of the Origin of Microseisms}, {\it Royal Society of London
  Philosophical Transactions Series A\/}, {\bf 243}, 1--35.

\bibitem[{Marquering} et~al.(1999){Marquering}, {Dahlen}, \& {Nolet}]{dahlen99}
{Marquering}, H., {Dahlen}, F.~A., \& {Nolet}, G., 1999.
\newblock {Three-dimensional sensitivity kernels for finite-frequency
  traveltimes: the banana-doughnut paradox}, {\it Geophysical Journal
  International\/}, {\bf 137}, 805--815.

\bibitem[{Nawa} et~al.(1998){Nawa}, {Suda}, {Fukao}, {Sato}, {Aoyama}, \&
  {Shibuya}]{nawa98}
{Nawa}, K., {Suda}, N., {Fukao}, Y., {Sato}, T., {Aoyama}, Y., \& {Shibuya},
  K., 1998.
\newblock {Incessant excitation of the Earth's free oscillations}, {\it Earth,
  Planets, and Space\/}, {\bf 50}, 3--8.

\bibitem[Pedersen et~al.(2007)Pedersen, Kr\"uger, \& the SVEKALAPKO Seismic
  Tomography Working~Group]{hellegji06}
Pedersen, H., Kr\"uger, F., \& the SVEKALAPKO Seismic Tomography Working~Group,
  2007.
\newblock Influence of the seismic noise characteristics on noise correlations
  in the {Baltic Shield}, {\it Geophysical Journal International\/}, {\bf 168},
  197--210.

\bibitem[{Prieto} et~al.(2009){Prieto}, {Lawrence}, \& {Beroza}]{prieto09}
{Prieto}, G.~A., {Lawrence}, J.~F., \& {Beroza}, G.~C., 2009.
\newblock {Anelastic Earth structure from the coherency of the ambient seismic
  field}, {\it Journal of Geophysical Research (Solid Earth)\/}, {\bf 114},
  7303.

\bibitem[{Prieto} et~al.(2011){Prieto}, {Denolle}, {Lawrence}, \&
  {Beroza}]{prieto11}
{Prieto}, G.~A., {Denolle}, M., {Lawrence}, J.~F., \& {Beroza}, G.~C., 2011.
\newblock {On amplitude information carried by the ambient seismic field}, {\it
  Comptes Rendus Geoscience\/}, {\bf 343}, 600--614.

\bibitem[{Rhie} \& {Romanowicz}(2004)]{rhie04}
{Rhie}, J. \& {Romanowicz}, B., 2004.
\newblock {Excitation of Earth's continuous free oscillations by
  atmosphere-ocean-seafloor coupling}, {\it \nat\/}, {\bf 431}, 552--556.

\bibitem[{Rivet} et~al.(2011){Rivet}, {Campillo}, {Shapiro}, {Cruz-Atienza},
  {Radiguet}, {Cotte}, \& {Kostoglodov}]{rivet11}
{Rivet}, D., {Campillo}, M., {Shapiro}, N.~M., {Cruz-Atienza}, V., {Radiguet},
  M., {Cotte}, N., \& {Kostoglodov}, V., 2011.
\newblock {Seismic evidence of nonlinear crustal deformation during a large
  slow slip event in Mexico}, {\it {Geophys. Res. Lett.}\/}, {\bf 38}, 8308.

\bibitem[{Roux} et~al.(2005){Roux}, {Sabra}, {Kuperman}, \& {Roux}]{roux05}
{Roux}, P., {Sabra}, K.~G., {Kuperman}, W.~A., \& {Roux}, A., 2005.
\newblock {Ambient noise cross correlation in free space: Theoretical
  approach}, {\it Acoustical Society of America Journal\/}, {\bf 117}, 79--84.

\bibitem[{Snieder}(2004)]{snieder04}
{Snieder}, R., 2004.
\newblock {Extracting the Green's function from the correlation of coda waves:
  A derivation based on stationary phase}, {\it \pre\/}, {\bf 69}(4), 046610.

\bibitem[{Stehly} et~al.(2006){Stehly}, {Campillo}, \& {Shapiro}]{stehly06}
{Stehly}, L., {Campillo}, M., \& {Shapiro}, N.~M., 2006.
\newblock {A study of the seismic noise from its long-range correlation
  properties}, {\it {Journal of Geophysical Research}\/}, {\bf 111}(B10306).

\bibitem[{Tromp} et~al.(2010){Tromp}, {Luo}, {Hanasoge}, \& {Peter}]{tromp10}
{Tromp}, J., {Luo}, Y., {Hanasoge}, S., \& {Peter}, D., 2010.
\newblock {Noise cross-correlation sensitivity kernels}, {\it Geophysical
  Journal International\/}, {\bf 183}, 791--819.

\bibitem[{Tsai}(2009)]{tsai09}
{Tsai}, V.~C., 2009.
\newblock {On establishing the accuracy of noise tomography travel-time
  measurements in a realistic medium}, {\it Geophysical Journal
  International\/}, {\bf 178}, 1555--1564.

\bibitem[{Tsai}(2010)]{tsai10}
{Tsai}, V.~C., 2010.
\newblock {The relationship between noise correlation and the Green's function
  in the presence of degeneracy and the absence of equipartition}, {\it
  Geophysical Journal International\/}, {\bf 182}, 1509--1514.

\bibitem[{Tsai}(2011)]{tsai11}
{Tsai}, V.~C., 2011.
\newblock {Understanding the amplitudes of noise correlation measurements},
  {\it Journal of Geophysical Research (Solid Earth)\/}, {\bf 116}, 9311.

\bibitem[Weaver et~al.(2009)Weaver, Froment, \& Campillo]{WeaverJASA09}
Weaver, R., Froment, B., \& Campillo, M., 2009.
\newblock On the correlation of non-isotropically distributed ballistic scalar
  diffuse waves, {\it J. Acoust. Soc. America\/}, {\bf 126}, 1817--1826.

\bibitem[{Weaver} et~al.(2011){Weaver}, {Hadziioannou}, {Larose}, \&
  {Campillo}]{weaver11}
{Weaver}, R.~L., {Hadziioannou}, C., {Larose}, E., \& {Campillo}, M., 2011.
\newblock {On the precision of noise correlation interferometry}, {\it
  Geophysical Journal International\/}, {\bf 185}, 1384--1392.

\bibitem[Wegler \& Sens-Schonfelder(2007)]{wegler07}
Wegler, U. \& Sens-Schonfelder, C., 2007.
\newblock Fault zone monitoring with passive image interferometry, {\it
  Geophysical Journal International\/}, {\bf 168}(3), 1029--1033.

\bibitem[{Woodard}(1997)]{woodard}
{Woodard}, M.~F., 1997.
\newblock {Implications of Localized, Acoustic Absorption for Heliotomographic
  Analysis of Sunspots}, {\it \apj\/}, {\bf 485}, 890.

\bibitem[{Wu} \& {Aki}(1985)]{wu85}
{Wu}, R.~S. \& {Aki}, K., 1985.
\newblock {Elastic wave scattering by a random medium and the small-scale
  inhomogeneities in the lithosphere}, {\it \jgr\/}, {\bf 90}, 10261--10274.

\bibitem[Yang \& Ritzwoller(2008)]{YangG308}
Yang, Y. \& Ritzwoller, M., 2008.
\newblock The characteristics of ambient seismic noise as as source for surface
  wave tomography, {\it Geochemistry, Geophysics, Geosystems\/}, {\bf 9}.

\bibitem[{Zaccarelli} et~al.(2011){Zaccarelli}, {Shapiro}, {Faenza}, {Soldati},
  \& {Michelini}]{shapiro11}
{Zaccarelli}, L., {Shapiro}, N.~M., {Faenza}, L., {Soldati}, G., \&
  {Michelini}, A., 2011.
\newblock {Variations of crustal elastic properties during the 2009 L'Aquila
  earthquake inferred from cross-correlations of ambient seismic noise}, {\it
  {Geophys. Res. Lett.}\/}, {\bf 38}, 24304.

\end{thebibliography}

\appendix
\section{Fourier Convention}\label{convention}
The following Fourier transform convention is utilized
\begin{eqnarray}
\int_{-\infty}^\infty dt~e^{i\omega t}~ g(t) &=& {\hat g}(\omega) ,\\
\int_{-\infty}^\infty dt~e^{i\omega t} &=& 2\pi~\delta(\omega),\label{inv.fourier}\\
\frac{1}{2\pi}\int_{-\infty}^{\infty} d\omega~e^{-i\omega t}~ {\hat g}(\omega) &=& g(t),\\
\int_{-\infty}^\infty d\omega~e^{-i\omega t} &=&  2\pi~\delta(t),
\end{eqnarray}
where $g(t), {\hat g}(\omega)$ are a Fourier-transform pair. 
The equivalence between cross-correlations and convolutions in the Fourier and temporal domain are written so
\begin{equation}
h(t) = \int_{-\infty}^{\infty} dt'~ f(t')~ g(t+t') \Longleftrightarrow {\hat h}(\omega) = {\hat f}^*(\omega)~{\hat g}(\omega),\label{cross.c} \\
\end{equation}
\begin{equation}
h(t) = \int_{-\infty}^{\infty} dt'~ f(t')~ g(t-t') \Longleftrightarrow {\hat h}(\omega) =  {\hat f}(\omega)~{\hat g}(\omega).
\end{equation}
The following relationship also holds (for real functions $f(t), g(t)$)
\begin{equation}
\int_{-\infty}^\infty dt~f(t)~g(t) = \frac{1}{2\pi}\int_{-\infty}^{\infty} d\omega~{\hat f}^*(\omega)~{\hat g}(\omega) = \frac{1}{2\pi}\int_{-\infty}^{\infty} d\omega~{\hat f}(\omega)~{\hat g}^*(\omega).
\end{equation}

\end{document}